\documentclass[10pt,a4paper,graphicx]{article}

\usepackage{amsmath}
\usepackage{amssymb}
\usepackage[mathscr]{eucal}
\usepackage{theorem}

\makeatletter
\renewenvironment{figure}
  {\let\@capwidth\linewidth\def\@captype{figure}}
  {}

\theorembodyfont{\upshape}
\newtheorem{theorem}{Theorem}[subsection]
\newtheorem{definition}[theorem]{Definition}
\newtheorem{corollary}[theorem]{Corollary}

\input{epsf}

\begin{document}

\vspace{1cm}
\begin{center}
\LARGE{ {\bf Coarse graining and control theory model reduction}}

\vspace{2cm}

\large{David E. Reynolds}~\footnote{Department of Physics, University of California,Santa Barbara, CA 93106}
\end{center}

\vspace{2cm}











\begin{quote}
{\bf ABSTRACT:} We explain a method, inspired by control theory model reduction
and interpolation theory, that rigorously establishes the types of 
coarse graining that are appropriate for  systems with 
quadratic, generalized Hamiltonians.  
For such systems, 
general conditions are given that 
establish when local coarse grainings should be valid.
Interestingly, our analysis provides a reduction method that is 
valid regardless of whether or not the system is isotropic.  We provide 
the linear harmonic chain as a prototypical example.  
Additionally, these reduction techniques are based on the dynamic response
of the system, and hence are also applicable to nonequilibrium systems.

\vspace{.2cm}

\end{quote}

\pagebreak

\section{Introduction}

Despite many of the great successes of statistical mechanics, 
it  still lacks adequate methods for systematically 
treating heterogeneous and nonequilibrium systems.  This is 
especially disconcerting considering that much of the world about us
is both heterogeneous and far from equilibrium. 
In addition, the treatment of open systems and systems with nontrivial boundary conditions
have yet to be systematically incorporated into statistical mechanics 
\footnote{Of course the latter concern is somewhat atypical considering that boundary terms are usually deemed unimportant in the thermodynamic limit.}.
It is not the purpose of this paper to address all of these deficiencies.
Rather, we introduce techniques from control theory engineering and 
interpolation theory to shed new light on such problems. 

It is a standard practice in physics to simplify complicated 
systems.
In particular limits, such as high-temperature or low-density,
these idealizations may become exact.
We will discuss two of the main methods used in physics to construct reduced-order models.

The projection-operator formalism (POF) of Mori and Zwanzig \cite{mori65a,mori65b,zwanzig73,hansen}
is a method from nonequilibrium statistical mechanics.  
It allows contact between the constitutive 
conservative microscopic equations and the more macroscopic 
phenomenological Langevin equations.
The key mathematical ingredient in this approach,
given an arbitrary observable, is to project along 
particular ``directions'' in state space in order to obtain 
an alternative evolution equation involving contributions from a forcing term
and from a memory kernel.  Here the projections involved are simply 
integrations over the appropriate phase space variables.  
A textbook application of the POF is a particle in a heat bath
\cite{zwanzig73,ford87,vankampen,barahona02}.
In this example, there is a clear split between important (system) variables and 
the less important (environment) variables.  
Thus, taking the system variables as the particle's position and momentum 
justifies projecting out the bath variables.

The renormalization group (RG) from field theory and
 equilibrium statistical mechanics \cite{gold,Lesne},
in its original form, involves identifying how the physics of a system changes with 
scale.  Equivalently, the renormalization group identifies how the parameters of a system's 
Hamiltonian or Lagrangian vary as the system is coarse grained.
In the RG,  systems are, almost invariably, locally coarse grained (i.e. locally-averaged).
In the context of equilibrium statistical mechanics, the coarse graining is realized 
with the appropriate partial trace of a Boltzmann weight.  
Formally, the partial trace is 
equivalent to the projections used in the POF.

An important observation is that both the POF and RG are completely general
techniques.  Although typical system reductions are either based on \emph{a priori} 
system-environment splits or obvious symmetries dictating local coarse graining, 
there is enormous ambiguity in choosing which states to trace out
\footnote{Taking the partial trace of the probability density (Boltzmann weight) produces
the probability density for the corresponding random variable.  Thus, the only real constraint 
one should put on the partial trace is that it corresponds to a measurable random variable.}.
Intuition is enough of a guide 
for determining how to coarse grain homogeneous systems with local interactions. However,
without some direction for dealing with heterogeneous systems, possibly with nonlocal interactions,
the POF and RG are too general; they are useless.
For instance, locally averaging about the interface in a layered system 
loses important information about the system. Additionally, locally averaging such  
systems is actually more likely to complicate the model.
Complications arise since the averaged 
theory would pick up extra couplings to enforce the constraint of well-defined boundaries
and induce couplings between the bulk of the different layers. 
In short, for general systems, local coarse graining is likely
to discard important details.  Consequently, as the effective influence of these discarded
details is reincorporated into the coarsened description of the system,  the new effective theory
becomes increasingly complicated.

The above considerations support the view advocated in the work by Bricmont and Kupiainen 
\cite{bricmont88,bricmont98,bricmont01}. They contend that systems should not be blindly coarse grained
scale by scale, but rather, large fluctuations should remain fixed while those 
degrees of freedom corresponding to small fluctuations are integrated away.  
A direct consequence of this perspective is that 
nonlocal coarse graining is on the same footing as its local counterpart. 
Intuitively, internal states that cause the largest fluctuations are the most relevant. 
The problem with such a program is that there exists no general framework that allows for 
an unambiguous measure of the relative importance of a system's internal degrees of freedom.

It is our claim that methods from control theory and modern interpolation theory
provide a complete, general framework for determining how to appropriately coarse grain
linear and linearly-dominated nonlinear systems.  Consequently, this opens up 
many new possible avenues to address the full nonlinear problem
\cite{scherpen93,scherpen96,scherpen00}.  The primary idea
of this approach is to coarse grain a system based on its dynamic response.  
For linear systems, it is possible to develop a completely unambiguous measure of 
how the internal states of a system contribute to the response.  In other words,
it is possible to assign a relative importance to the internal degrees of freedom.
Determining this measure then dictates how the system should be coarse 
grained. An especially nice feature of the control theory analysis is 
that it decomposes the response into two separate, physically intuitive, parts:  the 
controllability and observability of the internal states.  
Furthermore, these techniques are not limited to the idealized setting in which all of the internal 
degrees of freedom of the system can be perfectly measured.  In fact, these methods
were tailored to deal with physical systems in an experimental setting!
They are applicable even if 
the actuators and sensors interfaced with the system are imperfect.  
These methods are not only of great theoretical use; they are of practical 
use as well.
 
In 
work by Hartle and Brun \cite{brun99}, 
it is speculated
that local coarse graining produces more deterministic effective equations of motion 
than nonlocal coarse graining.  The problem with this claim is that it was made 
based on investigating the homogeneous linear harmonic chain on $\mathbb{Z}_N$ (i.e. on a ring) 
and considering a set of measure zero of all possible ways to coarse grain the system.
The main result of our paper rigorously establishes for what (linear) systems the above claim 
is true and how it breaks down for general linear systems.  A primary instance when it breaks 
down is for heterogeneous systems.  
We also establish how to appropriately coarse grain systems when 
local coarse graining breaks down.

This paper serves two functions; (1) to introduce and integrate basic concepts 
from control theory into standard physics problems, and (2) to develop and apply 
a new algorithm for coarse graining that complements existing physical reduction techniques. 
In Section \ref{sec:tutorial} we provide background material on the 
open loop control of linear systems. The definitions of controllability and observability 
are made precise.
The controllability and observability operators and gramians are then introduced. 
From these objects, we establish a simultaneous measure 
of controllability and observability.
This measure specifies the relative importance of different 
internal degrees of freedom.  
It also dictates how to model reduce or, equivalently, to coarse grain.
Appendix \ref{sec:infiniteH} contains important details that generalize the
control theory model reduction 
techniques in Section \ref{sec:tutorial} to conservative and unstable
systems. The lower bound in Appendix \ref{sec:infiniteH} is a new result.
Lastly, in Section \ref{sec:Ham}, we apply model reduction techniques
to oscillator 
systems to determine the ``natural'' reductions they admit.
We see that under some circumstances, depending  
on the spectral content of the system, local coarse 
graining is valid.  
We also show how
to coarse grain a system even if it is not homogeneous and isotropic.  Local coarse graining
cannot be expected to be appropriate for
general quadratic Hamiltonians.
In fact, our analysis shows the precise manner in which it is not. 
For illustrative purposes, we examine the linear harmonic chain in detail.


\section{A control theory tutorial}
\label{sec:tutorial}

This section describes how control theory 
methods, in particular Hankel norm analysis, may be used to 
determine the relative importance of the internal degrees of freedom for arbitrary linear systems.
A state's importance is directly related to its contribution to the system's response.
In this section, we introduce the requisite control theory 
terminology and notation that will be used throughout.

In the opening subsection, we introduce the linear systems under investigation, their 
corresponding input-output behavior (response), and some requisite material on the 
realization theory of input-output operators.  In the next subsection, we provide definitions and 
measures of controllability and observability.  The Hankel operator, 
its interpretation in terms of controllability and observability, and its 
relation to balanced realizations 
comprise the final subsection.
The latter topics are especially important in control theory model reduction 
and, consequently, also for coarse graining.  
Although we made
no attempt for this tutorial to be an exhaustive review, 
we include enough detail for the paper to be self-contained. 
All theorems in this section are stated without proof.
The interested reader is encouraged to consult the following references
\cite{robust1,robust2,glover84,feintuch,vinnicombe}.  Those who are already 
familiar with the above concepts 
may comfortably skip ahead to Section \ref{sec:Ham}.

\subsection{Linear Systems and Realizations}
\label{subsec:realizations}

\begin{figure}
\centerline{\epsfxsize=6cm \epsfbox{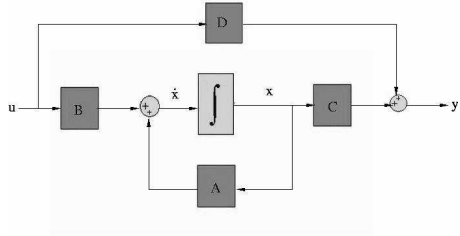}}
\caption{ {\small A block diagram representation of the linear system from \ref{eq:lin1} and \ref{eq:lin_sol}.
Directed lines flowing into boxes represents vectors being multiplied by operators (or matrices).
For instance,  initially $u$ flows into $\mathbf{B}$, hence the output of the first box is $\mathbf{B}u$.
The circles in the diagram are adders. Vectors that flow into adders are summed.}}
\label{fig:block_diagram}
\end{figure}

This paper concerns 
linear time invariant
(LTI) systems (i.e. linear systems with time translation invariance) of the form:
\begin{equation}
\label{eq:lin1}
\begin{array}{c}

\dot{x} = \mathbf{A}x + \mathbf{B}u  \\
y = \mathbf{C}x + \mathbf{D}u	  
\end{array}, \ \ \ 
\left\{ 
\begin{array}{l}
t \geq t_{0}, \\
x(t) \in \mathbb{R}^{n}, \\
u(t) \in \mathbb{R}^{m}, \\
y(t) \in \mathbb{R}^{p},
\end{array}
\right.
\end{equation}
where $x$ is the ``internal'' state of the system, $y$ represents quantities directly
measured by appropriately positioned sensors, and $u$ represents the external driving force.  
The matrix $\mathbf{A}$ captures the natural dynamics of the 
system, while, respectively, $\mathbf{B}$ and $\mathbf{C}$ dictate which internal states of the system are in 
contact with the driving and measurement.  $\mathbf{D}$ is responsible for the
feedthrough of the system.  Feedthough is the (possibly amplified) contribution of the driving that 
is directly measured.  A control theoretic, diagram representing  this system 
is in Figure \ref{fig:block_diagram}. 
Expressing the system in this form reflects that only partial information is measured
and that the system is open.
Since the system is LTI,
$\mathbf{A}$, $\mathbf{B}$, $\mathbf{C}$, and $\mathbf{D}$ are constant coefficient matrices.

The general solution to the above problem is given by:
\begin{equation}
\label{eq:lin_sol}
y(t) = \underbrace{\mathbf{C}e^{\mathbf{A} (t-t_0)}x_0}_{\textrm{zero input response}} \quad+\quad 
\underbrace{\int_{t_0}^{t} \mathbf{C}e^{\mathbf{A} (t-\tau)}
\mathbf{B}u(\tau)\mathrm{d}\tau}_{\textrm{zero state response}} 
\quad+\quad \underbrace{\mathbf{D}u(t)}_{\textrm{feedthrough}}
\end{equation}  
If we consider only the zero state response (trivial initial conditions), then  
\begin{equation}
\label{eq:i_o_operator}
\begin{array}{c}
y = Gu  \\ \ = \int \mathcal{K}(t,\tau)u(\tau) \mathrm{d}\tau = 
\int_{t_0}^{t} \mathbf{C}e^{\mathbf{A(t-\tau)}}\mathbf{B}u(\tau) \mathrm{d}\tau.
\end{array}
\end{equation} 
The integral kernel has many names. In the time domain it is referred to as the impulse 
response or the Green's function.  Alternatively, for stable, LTI systems, 
the Fourier transform of the integral kernel is also known as the transfer matrix, 
Green's function in the frequency domain, or the propagator.
Since feedthrough is not crucial in our analysis, $\mathbf{D}=0$
from this point forward.

In control theory, a system is specified by an experiment. This is reflected 
by the dependence of $G$ on $\mathbf{B}$,$\mathbf{C}$, and $\mathbf{D}$.  
A system is defined by its response (i.e. by $G$).  From an experiment, the 
only available data is from the inputs and outputs.  
Hence, the matrices $(\mathbf{A},\mathbf{B},\mathbf{C},\mathbf{D})$ are unknown.
Constructing all of such matrices corresponding to a given 
response is the objective of realization theory. 
The system matrices $(\mathbf{A},\mathbf{B},\mathbf{C},\mathbf{D})$
form a state space realization of the system.
For a given system, there does not exist a unique realization.  However, given a realization, 
there exists a unique system.
The choice of experiment fixes the system's inputs and outputs. 
The remaining ambiguity is due to the internal states of the system.
For instance, an arbitrary invertible, linear change of variables, 
$x=\mathbf{R}z$, demonstrates this.
$(\mathbf{R}^{-1}\mathbf{A}\mathbf{R},\mathbf{R}^{-1}\mathbf{B},\mathbf{C}\mathbf{R},\mathbf{D})$
is also a realization of $G$. Since $G$ is invariant under the above similarity transformations, 
realizations belong to equivalence classes.  The remaining indeterminateness arises 
since there is not a bound on the number of internal degrees of freedom.  In fact, there may be arbitrarily
many internal states that do not contribute to the system's response.  
State space realizations that have the minimal internal state dimension are called 
minimal realizations \footnote{Minimal realizations of a given state dimension all belong to the same equivalence class.}.
Although $G$ is typically an infinite rank operator
(i.e. the image of $G$ is infinite dimensional), 
the internal state dimension of its minimal 
realizations gives it an order. 
When the minimal realization is stable, the order of $G$ is also known as its McMillan degree.  

Minimal realizations represent the part of the system that is
observable and controllable.
To clarify this, we will introduce the definitions of controllability and 
observability.  
We then define  the controllability and oberservability operators along with their respective gramians.
These  concepts are imperative to  assigning a measure of how much an individual 
internal state contributes to the system's response.


\subsection{Controllability and Observability}
\label{sec:c_and_o}
$\bullet$ {\bf Controllability}

{\it Controllability concerns the effects of driving on the system.  In particular,
a system is controllable if it is possible to drive a system from any 
initial configuration to any final configuration.  An internal state of a system is considered 
uncontrollable if it cannot be driven to every other state.}

The issue of whether or not a system (or state) is controllable is a yes-or-no question.
However, we may still intuitively assign a degree of controllability to a state. 
An example of this is to consider the response of a conservative system when it is 
driven at one of its characteristic frequencies (at resonance).  This is mathematically 
realized by a divergence (or a peak, in general) in the Fourier transform of the response.  
This is the simplest example of a system's mode being very controllable, insofar that we can
elicit a large response from small amplitude 
driving\footnote{Driving with small amplitude is also termed driving or forcing with small gain.}. 
It is easy to drive states in the direction of this mode.
Generally input-output resonances do not always correspond to internal resonances.
Should there be a set or subspace of state or phase space that can not be reached 
via driving, then such ``directions'' are uncontrollable.
A system is controllable if every direction in state or phase space is controllable.
The following provides a more formal and precise definition of controllability.

\begin{definition} 
\label{cdef}
A system is controllable if it is possible
to drive any initial  state $x_0$ to any final state $x_f$ in any nonzero time interval.
\end{definition}

$\bullet$ {\bf Observability}

{\it  Observability describes how easily the internal state of the system
can be reconstructed from measurements of the output.  Intimately connected 
to this is the precise determination of the internal 
initial conditions.  Initial conditions that cannot be reconstructed are the 
system's unobservable states. }

The mechanical model of a particle in a heat bath provides a physical example of observable 
versus unobservable states.  As alluded to earlier, such as system admits a natural 
system-environment split.
In this case the system is the single oscillator, while the environment is the bath.  
The oscillator is the primary object under investigation and hence, an experimental apparatus is devised to 
measure its displacement and/or velocity.  Since the bath is composed of 
innumerable constituent particles (or oscillators), the individual trajectories of the bath
particles are unknown.  
While the single oscillator is strongly observable, the bath is only weakly observable.
It is possible to reconstruct the initial conditions for the single oscillator
but not for the entire bath. A more precise and formal definition of observability is:

\begin{definition} 
A system is observable if it is possible
to fully determine any initial state $x_0$ by measuring y over
 any nonzero time interval.
\end{definition}

$\bullet$ {\bf Controllability and observability operators}

 It is clear that the input-output operator, $G$, from equation \eqref{eq:i_o_operator}
takes in inputs from $u$ from some space and outputs $y$ in another.
For concreteness, from now on we will 
consider the domain of $G$ to be $\mathcal{L}^m_2[-T,T]$ (i.e. $m$ copies of $\mathcal{L}_2$)
and the range to be $\mathcal{L}^p_2[-T,T]$. More generally, $u$ may also 
be a vector in $\mathcal{L}^m_1$,
 $\mathcal{L}^m_{\infty}$, or a Langevin contribution to the dynamics.
The construction of the controllability and observability operators, $\Psi_c$ and $\Psi_o$ respectively, 
is largely motivated by the fact that the Hankel
operator, to be introduced later, can be factored into their product.
Thus, the response
 may be 
decomposed into observability and controllability.  

The controllability operator is defined by:
\begin{equation}
\label{eq:psi_c}
\begin{array}{c}
\Psi_c:\mathcal{L}^m_2[-T,0] \to \mathbb{R}^n \\
\Psi_c u = \int_{-T}^{0} e^{-\mathbf{A}\tau}\mathbf{B}u(\tau) \mathrm{d}\tau
=\int_{0}^{T} e^{\mathbf{A}\tau}\mathbf{B}u(-\tau) \mathrm{d}\tau
\end{array}
\end{equation} 
Formally $\Psi_c$ is not defined on the full domain of $G$.  It can be extended to the full space, however, 
by  defining $\mathcal{L}^m_2[0,T]$ to be in its null space.
The controllability operator allows for an algebraic definition of controllability.
\begin{theorem} 
A linear system, as in equation \eqref{eq:lin1}, specified by 
$(\mathbf{A},\mathbf{B},\mathbf{C},\mathbf{D})$
is controllable if and only if 
the image of $\Psi_c$
 is all of  $\mathbb{R}^{n}$.
\end{theorem}
If a system in equation \eqref{eq:lin1} is controllable, 
we call the system pair $(\mathbf{A},\mathbf{B})$
controllable.  Additionally, the space of states that are controllable 
forms an $\mathbf{A}$-invariant subspace.  The controllable 
subspace is precisely the image of $\Psi_c$,$\ \mathcal{R}(\Psi_c)$.

Similarly, the observability operator is defined by:
\begin{equation}
\label{eq:psi_o}
\begin{array}{c}
\Psi_o:\mathbb{R}^n \to \mathcal{L}^p_2[0,T] \\
\Psi_o z = \mathbf{C} e^{\mathbf{A}t}z
\end{array}
\end{equation} 
In contrast to controllability, the set of observable states do not 
form an invariant subspace.  The span of the unobservable 
states forms an $\mathbf{A}$-invariant subspace\footnote{ This and 
its controllable analog are important because they are responsible for the 
Kalman decomposition.}.
The null space of $\Psi_o$, $\ Null(\Psi_o)$, comprises the unobservable subspace.  
This implies that a system is observable 
provided that $\Psi_o$ has full rank (i.e. the null space is empty).
This gives the formal algebraic definition of observability.
\begin{theorem}[Test for observability]
\label{obs_operator1}
A linear system given by \eqref{eq:lin1} is observable iff $\mathrm{rank}(\Psi_o)=n$
(i.e. $\Psi_o$ has full rank).
\end{theorem}
If a system is observable,
we call the system pair $(\mathbf{C},\mathbf{A})$ observable. 
Controllability and 
observability are completely dual to each other.  For example,
  $(\mathbf{A},\mathbf{B})$ is controllable if and only if 
 $(\mathbf{B}^{\dagger},\mathbf{A}^{\dagger})$ is observable.

$\bullet$ {\bf Controllability and observability gramians}

Superficially it may seem that the above operators only give us limited information.
Specifically, we only have binary tests for controllability and observability
based on whether or not the state is in $\mathcal{R}(\Psi_c)$ or in $Null(\Psi_o)$.
Our objective is to determine
how observable and controllable a state is in order to
quantify its contribution to the response.  It is precisely the controllability 
and observability gramians that provide this information.  However, as 
will soon become apparent, the operators are intimately related to the gramians.

Determining $\mathcal{R}(\Psi_c)$ and $Null(\Psi_o)$ is a 
formidable challenge since the domain of $\Psi_c$ and the range of $\Psi_o$ are
infinite dimensional spaces.  However, the formal
operator adjoint 
makes the problem more tractable.
  Since $\Psi_c:\mathcal{L}_2^m[-T,0] \to \mathbb{R}^n$
this then implies that $\Psi_c^{\dagger}:\mathbb{R}^n \to \mathcal{L}^m_2[-T,0]$, 
where $\Psi_c^{\dagger}$ is the operator adjoint of $\Psi_c$. 
Similarly, $\Psi_o^{\dagger}: \mathcal{L}_2^p[0,T] \to \mathbb{R}^n $.  
An important property of the adjoint of an operator is that its image
is perpendicular to the original operator's null space, that is 
$\mathcal{R}(T^{\dagger}) \bot Null(T)$.  
Also $\mathcal{R}(\Psi_c^{\dagger}) \bot Null(\Psi_c)$ and 
 $Null(\Psi_o) \bot \mathcal{R}(\Psi_o^{\dagger})$.
Consequently, $\mathcal{R}(\Psi_c) = \mathcal{R}(\Psi_c\Psi_c^{\dagger})$
and  $Null(\Psi_o) = Null(\Psi_o^{\dagger}\Psi_o)$.  However, since
$\Psi_c^{\dagger}$ maps $\mathbb{R}^n$ to $\mathcal{L}^m_2[-T,0]$
and $\Psi_o^{\dagger}$ maps $\mathcal{L}^p_2[0,T]$ to $\mathbb{R}^n$, 
$\Psi_c\Psi_c^{\dagger}$ and $\Psi_o^{\dagger}\Psi_o$ are $n \times n$ matrices.  
Finding the controllability and observability subspaces reduces
to discovering the images and null spaces of $n \times n$ matrices.  

From the definition of the adjoint, the expression for
$\Psi_c^{\dagger}$ is
\begin{equation}
\label{eq:adj_psi_c}
\begin{array}{c}
\Psi_c^{\dagger}: \mathbb{R}^n \to \mathcal{L}^m_2[-T,0],\\
\Psi_c^{\dagger} z = \mathbf{B}^{\dagger} e^{-\mathbf{A}^{\dagger}t}z \quad 
\textrm{For all} \ z  \in \mathbb{R}^n, \ t \in [-T,0].
\end{array}
\end{equation} 
The following relation holds for  $\Psi_o^{\dagger}$:
\begin{equation}
\label{eq:adj_psi_o}
\begin{array}{c}
\Psi_o^{\dagger}:\mathcal{L}^p_2[0,T] \to \mathbb{R}^n, \\
\Psi_o^{\dagger} f = \int_{0}^{T} e^{\mathbf{A}^{\dagger}\tau}\mathbf{C}^{\dagger}f(\tau) \mathrm{d}\tau.
\end{array}
\end{equation}
From above, we are left with objects that are of fundamental importance
for establishing quantitative measures of controllability and observability, the gramians.  
The controllability gramian, $\mathbf{W}_c$, is defined by:
\begin{equation}
\label{eq:cgramian}
\begin{array}{c}
\mathbf{W}_c(T) = \Psi_c \Psi_c^{\dagger} \\
=\int_{0}^{T} e^{\mathbf{A}t}\mathbf{B}\mathbf{B}^{\dagger} e^{\mathbf{A}^{\dagger}t} \mathrm{d}t.
\end{array}
\end{equation}
The observability gramian, $\mathbf{W}_o$, is defined by:
\begin{equation}
\label{eq:ogramian}
\begin{array}{c}
\mathbf{W}_o(T) = \Psi_o^{\dagger} \Psi_o \\
=\int_{0}^{T} e^{\mathbf{A}^{\dagger}t}\mathbf{C}^{\dagger} \mathbf{C}e^{\mathbf{A}t} \mathrm{d}t.
\end{array}
\end{equation}
From their definitions, the gramians are both self-adjoint and positive semi-definite.
Additionally, the controllable subspace of the system is the image of 
$\mathbf{W}_c(T)$.  Thus, a linear system is controllable if and only if $\mathbf{W}_c(T)$ is 
nonsingular (invertible).  Similarly, the unobservability subspace is the null space 
of $\mathbf{W}_o(T)$.  A linear system is then observable if and only if  $\mathbf{W}_o(T)$ is 
nonsingular (i.e. the null space is empty).  Consequently, controllability and 
observability are determined by calculating two matrices,  $\mathbf{W}_c(T)$ and  $\mathbf{W}_o(T)$.
Equations \eqref{eq:cgramian} and \eqref{eq:ogramian} are computationally
not very useful.
It is typically easier to determine the gramians from the equations that 
they satisfy.
\begin{eqnarray}
\label{eq:diff_lyap1}
\frac{\mathrm{d}\mathbf{W}_c}{\mathrm{d}T} = \mathbf{A}\mathbf{W}_c+\mathbf{W}_c\mathbf{A}^{\dagger}
+\mathbf{B}\mathbf{B}^{\dagger};  \ \  \mathbf{W}_c(0) = 0
\\
\label{eq:diff_lyap2}
\frac{\mathrm{d}\mathbf{W}_o}{\mathrm{d}T} = \mathbf{A}^{\dagger}\mathbf{W}_o+\mathbf{W}_o\mathbf{A}
+\mathbf{C}^{\dagger}\mathbf{C}; \ \  \mathbf{W}_o(0) = 0
\end{eqnarray}
For stable systems, in the limit as $T \to \infty$, the gramians satisfy algebraic Lyapunov equations
\footnote{The Lyapunov equations are $\mathbf{A}\mathbf{W}_c+\mathbf{W}_c\mathbf{A}^{\dagger}
+\mathbf{B}\mathbf{B}^{\dagger}=0$ and $\mathbf{A}^{\dagger}\mathbf{W}_o+\mathbf{W}_o\mathbf{A}
+\mathbf{C}^{\dagger}\mathbf{C}=0$.}.

\begin{figure}
\centerline{\epsfxsize=6cm \epsfbox{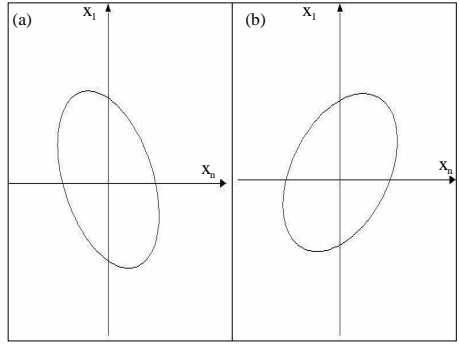}}
\caption{ { \small (a) depicts a controllability ellipsoid, while (b) depicts an observability
ellipsoid.  The semimajor axis of the ellipsoid in (a) indicates the most controllable  
direction in state space, and for the ellipsoid in (b) it indicates the most observable direction.}}
\label{fig:gramians}
\end{figure}

The directions in state or phase space corresponding to trivial eigenvalues of $\mathbf{W}_c$
are uncontrollable.  
Therefore, the eigenvectors of $\mathbf{W}_c$ corresponding to small eigenvalues
are only weakly controllable. It is along those directions that 
the controllability gramian is almost singular.  
Physically,
it requires much higher gains to reach these states than the more controllable states.
More precisely, consider the quadratic energy functional
 $F(u) = \int_{-T}^{0} \| u(t) \|_{\mathbb{R}^m}^2 \mathrm{d}t = \|u\|^2_{\mathcal{L}^m_2[-T,0]}$
 that measures the energy due to driving.
The $u$ that expends the least energy to reach a state $\bar{x} \in R^n$ from the origin 
\footnote{This is provided that the system is controllable
and the dynamics of the system are restricted to satisfy
\eqref{eq:lin1} with $\mathbf{D}=0$.}
is given by $u_{min} = \Psi_c^{\dagger}\mathbf{W}_c^{-1}\bar{x}$.  The energy due to such
driving is
\begin{equation}
\label{eq:min_energy}
\begin{array}{l}
 \|u_{min}\|^2_{\mathcal{L}^m_2[-T,0]} = \big<u_{min},u_{min}\big>_{\mathcal{L}^m_2} \\ \ 
= \big<  \Psi_c^{\dagger}\mathbf{W}_c^{-1}\bar{x}, \Psi_c^{\dagger}\mathbf{W}_c^{-1}\bar{x}\big>_{\mathcal{L}^m_2} \\ \  
= \big<\bar{x},\mathbf{W}_c^{-1}\Psi_c \Psi_c^{\dagger}\mathbf{W}_c^{-1}\bar{x}\big>_{\mathbb{R}^n} 
= \big<\bar{x},\mathbf{W}_c^{-1}\bar{x}\big>_{\mathbb{R}^n} \\ \  
= \|\mathbf{W}_c^{-1/2}\bar{x}\|_{\mathbb{R}^n}^2.
\end{array}
\end{equation}
If we drive the system in state or phase space with minimal 
force $ \|u_{min}\|^2 \le 1$, the corresponding region in state or phase space 
is a solid ellipsoid in $\mathbb{R}^n$.  This set, depicted in Figure \ref{fig:gramians}(a), 
corresponds to  
$\{\bar{x}\in\mathbb{R}^n:\bar{x}^{\dagger}\mathbf{W}_c^{-1}\bar{x} \le 1\}$.  
This ellipsoid is also specified by 
$\{\bar{x}\in\mathbb{R}^n:\bar{x} = \mathbf{W}_c^{1/2}z, \|z\|_{\mathbb{R}^n} \le 1\}$.
While  $\|\mathbf{W}_c^{-1/2}\bar{x}\|^2$ measures 
energy expenditure,  
$\frac{\|\mathbf{W}_c^{-1/2}x\|^2}{\|x\|^2} = \frac{x^{\dagger}\mathbf{W}_cx}{x^{\dagger}x}$  
measures a state's controllability.  This confirms the intuition
that states corresponding to small eigenvalues of $\mathbf{W}_c$
are the least controllable.

Physically, the oberservability operator produces a response given an initial condition.
What does this reveal about the inverse problem of reconstructing the initial conditions from 
measurements of the system's response?  
Mathematically this problem is posed as determining 
\begin{displaymath}
\min_{\bar{x}\in \mathbb{R}^n}
\|y-\Psi_o\bar{x}\|^2_{\mathcal{L}^p_2[0,T]}.
\end{displaymath}
When $y$ is in the image of $\Psi_o$, it is possible to precisely specify the
initial conditions, $\bar{x}$.
Otherwise, the initial condition minimizing $\|y-\Psi_o\bar{x}\|^2_{\mathcal{L}^p_2[0,T]}$, given
an arbitrary  $y \in \mathcal{L}^p_2[0,T] $, is $\bar{x}_{opt} = \mathbf{W}_o^{-1}\Psi_o^{\dagger}y$.
In order to obtain a quantitative measure of observability, we need only
consider the outputs, $y = \Psi_o \bar{x}$.  Immediately we recognize that 
$\|y\|^2_{\mathcal{L}^p_2[0,T]}/\|\bar{x}\|^2_{\mathbb{R}^n} =
 \|\mathbf{W}_o^{1/2}\bar{x}\|^2_{\mathbb{R}^n}\big|_{\|\bar{x}\| \le 1}$
measures a state's observability.
For instance, initial conditions, $\bar{x}$, corresponding to  
small eigenvalues of $\mathbf{W}_o$ elicit smaller responses than other states
and, consequently, are less observable.  If noise is present, responses resulting  
from such initial conditions would not be observed at all.
The set $\{\bar{x}\in\mathbb{R}^n:\bar{x} = \mathbf{W}_c^{1/2}z, \|z\|_{\mathbb{R}^n} \le 1\}$
corresponds to the observability ellipsoid depicted in Figure \ref{fig:gramians}(b).  
The directions along which the the ellipsoid is long are the  
most observable.

The utility in considering controllability and observability separately
is that they have precise and experimentally relevant interpretations. 
A problem with this approach is that it initially obstructs the path
to ascribing measures of response to physical states.  For instance,
it is possible to model reduce based on either controllability or observability
\footnote{It has been shown in \cite{rowley03} that reductions based on the 
proper orthogonal decomposition (POD) is essentially equivalent to model 
reducing based on controllability.}.  Unfortunately, the measures of 
controllability and observability are not unique.
This is transparent after considering how $\mathbf{W}_o$ and  $\mathbf{W}_c$
transform under similarity transformations to the system.  $\mathbf{W}_o$ transforms as
 $\mathbf{W}_o \stackrel{\mathbf{R}}{\longrightarrow} 
\tilde{\mathbf{W}}_o=\mathbf{R}^{\dagger}\mathbf{W}_o\mathbf{R}$, while 
 $\mathbf{W}_c$ transforms as $\mathbf{W}_c \stackrel{\mathbf{R}}{\longrightarrow} 
\tilde{\mathbf{W}}_c=\mathbf{R}^{-1}\mathbf{W}_c(\mathbf{R}^{\dagger})^{-1}$.
Thus, the gramians are not invariant under an arbitrary linear coordinate 
transformation.  
However, $\mathbf{W}_c\mathbf{W}_o$ transforms as 
$\mathbf{W}_c\mathbf{W}_o \stackrel{\mathbf{R}}{\longrightarrow} 
\tilde{\mathbf{W}}_c\tilde{\mathbf{W}_o}=\mathbf{R}^{-1}\mathbf{W}_c\mathbf{W}_o\mathbf{R}$ under 
a similarity transformation of the system.  The eigenvalues of  $\mathbf{W}_c\mathbf{W}_o$
are invariants of the system, are intimately related to the Hankel operator of the system,
 and thus will prove invaluable for producing reduced-order models.


\subsection{The Hankel Operator, Balanced Realizations, and Model Reduction}
\label{subsec:MR}

The theory of model reduction
is closely related to that of system realizations.  In model reduction, the 
goal is to find realizations (i.e. the matrices $(\mathbf{A},\mathbf{B},\mathbf{C},\mathbf{D})$)
with minimal state dimension that approximately capture the system's input-output
characteristics.
Often being \lq\lq near'' the original system is enough to dramatically 
reduce the number of internal states needed to model the system.  
Here distance or \lq\lq nearness'' is defined by 
the standard induced-operator norm.
For example, given an operator $S$ that acts on $\mathcal{L}_2$, the induced norm takes the form 
$\|S\|_{\mathcal{L}_2,i} = \sup_{\|v\|_{\mathcal{L}_2}\le1} \|Sv\|_{\mathcal{L}_2}$, where 
$v$ is a vector in $\mathcal{L}_2$.
Also, supposing that $\tilde{S}$
approximates $S$,  we will be considering two measures of the error, 
the absolute error $\|S-\tilde{S}\|_{\mathcal{L}_2,i}$ and
 the relative error  $\|S-\tilde{S}\|_{\mathcal{L}_2,i}/\|S\|_{\mathcal{L}_2,i}$.
The relative error is more appropriate for unbounded operators,
since most approximations are asymptotic estimates.  
The relative error is also the noise-to-signal ratio.

It is useful to note that there is an explicit expression for the induced $\mathcal{L}_2$
norm for bounded 
(stable), LTI, causal operators.
Such operators are in the space $\mathcal{H}_{\infty}$.  
The primary difficulty with this formula is that 
it is difficult to use both numerically and analytically.  
To motivate the formula, recall that an 
arbitrary $m \times n$ matrix $\mathbf{M}$ can be decomposed as 
$\mathbf{M} = \mathbf{U}\mathbf{\Sigma}\mathbf{V}^{\dagger}$ (i.e. the singular value decomposition)
where $\mathbf{U}$ and $\mathbf{V}$ are respectively $m \times m$ and $n \times n$ unitary matrices and 
$\mathbf{\Sigma}$ is only nontrivial along the diagonal.  The diagonal values 
$\mathbf{\Sigma}_{ii} = \sigma_i(\mathbf{M}) \ge 0$ are called the singular values of $\mathbf{M}$.  
If $m>n$ then the singular values are the eigenvalues of $\sqrt{\mathbf{M}^{\dagger}\mathbf{M}}$,
otherwise they are the eigenvalues of $\sqrt{\mathbf{M}\mathbf{M}^{\dagger}}$.  
Here $\|\mathbf{M}\|_{\mathbb{C}^m,i} = \sigma_{\max}(\mathbf{M})$ where 
$\sigma_{\max}(\mathbf{M}) = \max_j \sigma_j(\mathbf{M})$ (i.e. the largest singular value).
For a bounded, LTI, causal 
operator $S$ such that $S(u) = \int_{-\infty}^{t} \mathbf{\mathcal{K}}_S(t-\tau)u(\tau) \mathrm{d}\tau$, 
$\|S\|_{\mathcal{L}_2,i} = \sup_{\omega \in \mathbb{R}} \sigma_{\max}(\hat{\mathbf{\mathcal{K}}}_S(\omega))$
where $\hat{\mathbf{\mathcal{K}}}_S(\omega)$ is the Fourier transform of $\mathbf{\mathcal{K}}_S(t)$.

Coarse graining and model reduction are intimately related.
While both are reduction methods,
coarse graining emphasizes the spatial nature of reductions.
Model reduction, as it will be presented here, emphasizes 
input-output resonances and approximating the response.
Our approach to coarse graining
is to identify the best way to model reduce, based on the response, and 
then ascertain the spatial structure of the reduction. 
The latter topic is elaborated upon for linear oscillator systems in the next section.
Unfortunately, the former issue is 
a mathematically unresolved problem.
The control/interpolation theoretic statement of this
problem for stable systems (in the induced $\mathcal{L}_2$-norm) 
is called the $\mathcal{H}_{\infty}$ model reduction problem.
\begin{definition}[$\mathcal{H}_{\infty}$ Model Reduction Problem]
Given a bounded, LTI, causal operator $G$ of McMillan degree $n$, such that $G:\mathcal{L}_2 \to \mathcal{L}_2$,
find $\inf_{\mathrm{deg}(\tilde{G})\le k} \| G - \tilde{G}\|_{\mathcal{L}_2,i}$ for 
$\tilde{G}$ a bounded, LTI, causal operator and $k<n$.
\end{definition}
Fortunately, enough is known about a related problem, the Hankel norm model reduction problem, 
to provide error bounds
to the above $\mathcal{H}_{\infty}$ problem.  By using results from Hankel operator 
analysis and Hankel norm model reduction, we will be able to deduce physically-important 
results about coarse graining. 

$\bullet$ {\bf The Hankel operator}

The input-output picture
corresponds to an experimental situation, albeit a complicated one.  
The full input-output operator, $G$, from $y=Gu$ 
represents continuously driving and measuring a system.  
This operator is difficult to study because
it does not separate observation from driving.  As will become apparent, the Hankel
operator is the part of the input-output operator where the operations of measurement
and forcing are separated.

To facilitate analysis, it is convenient to decompose $\mathcal{L}_2[-T,T]$ into 
$\mathcal{L}_2[-T,0] \oplus \mathcal{L}_2[0,T]$, in other words, a causal (analytic) decomposition.
LTI causal systems can be visualized in the following way:
\begin{equation}
\label{eq:G_decomp}
\begin{array}{c}
\left[
\begin{array}{c} 
y_- \\ y_+ 
\end{array}
\right] = 
\left[
\begin{array}{cc}
G_{11} & G_{12} \\ G_{21} & G_{22}
\end{array}
\right]
\left[
\begin{array}{c}
u_- \\ u_+
\end{array}
\right]
= 
\left[
\begin{array}{cc}
\mathcal{T}_G & 0 \\ \Gamma_G & \widetilde{\mathcal{T}}_G
\end{array}
\right]
\left[
\begin{array}{c}
u_- \\ u_+
\end{array}
\right],  \\
\left\{ 
\begin{array}{l}
y_+ \in \mathcal{L}_2^p[0,T], \\
y_- \in \mathcal{L}_2^p[-T,0], \\
u_+ \in \mathcal{L}_2^m[0,T], \\
u_- \in \mathcal{L}_2^m[-T,0]. 
\end{array}
\right.
\end{array}
\end{equation}
Causality implies that $G_{12} = 0$.  The additional constraint that the system is
LTI implies that  $\mathbf{\mathcal{K}}(t,\tau)$ is purely a function of $t-\tau$.
$\mathcal{T}_G$ and $\widetilde{\mathcal{T}}_G$ are Toeplitz operators, while 
$\Gamma_G$ is a Hankel operator.  
It is not vital nor required for the reader to be familiar with Hankel or Toeplitz operators.
The interested reader is encouraged to consult 
\cite{toeplitz1,toeplitz2,toeplitz3,peller,nikolski,partington,glover84,feintuch} to learn
more about these operators.
If we denote the projection operator onto $\mathcal{L}_2[0,T]$
by $P_+$, then $P_+^2 = P_+$ and $\Gamma_G = P_+ G\big|_{\mathcal{L}_2[-T,0]}$.
Since such projection operators can never increase the norm, it follows that
$\|\Gamma_G\| \le \|G\|$.  Similarly, 
$\|\mathcal{T}_G\| =\|\widetilde{\mathcal{T}}_G\| \le \|G\|$, where the first equality arises since 
 $\mathcal{T}_G$ and $\widetilde{\mathcal{T}}_G$ differ only by time reversal. 
 A somewhat surprising  fact is that $\|\mathcal{T}_G\| = \|G\|$ \cite{toeplitz1}.
Model reduction based on  $\mathcal{T}_G$ is equivalent 
to the full $\mathcal{H}_{\infty}$ problem for stable systems.
Unfortunately, 
the experiments represented by  $\mathcal{T}_G$ involve simultaneously driving and observing.

The Hankel operator, $\Gamma_G$, accepts inputs 
driving the internal state to some $x_0$ at time $t=0$.
Subsequently, the system is measured as it evolves in time.  The 
separation of driving and measurement allows for $\Gamma_G$ to be factored in the particularly
convenient way:
\begin{equation}
\label{eq:hankel_op1}
\begin{array}{l}
\Gamma_G(u) =  P_+ G\big|_{\mathcal{L}_2[-T,0]}u = 
P_+\int_{-T}^0 \mathbf{C}e^{\mathbf{A} (t-\tau)}\mathbf{B}u(\tau)\mathrm{d}\tau \\ \
= P_+\mathbf{C}e^{\mathbf{A}t} \int_{-T}^0 e^{-\mathbf{A}\tau}\mathbf{B}u(\tau)\mathrm{d}\tau
= \Psi_o \Psi_c u.
\end{array}
\end{equation}
The Hankel operator may be factored as a product of the observability operator and controllability 
operator: $\Gamma_G =  \Psi_o \Psi_c$.  It follows that
\begin{equation}
\label{eq:hankel_norm}
\begin{array}{l}
\|\Gamma_G\|_{\mathcal{L}_2,i}^2 = \|\Gamma_G^{\dagger}\Gamma_G\|_{\mathcal{L}_2,i} \\ \
= \|\Psi_c^{\dagger}\Psi_o^{\dagger}\Psi_o\Psi_c\|_{\mathcal{L}_2,i} 
= \|\Psi_o^{\dagger}\Psi_o\Psi_c\Psi_c^{\dagger}\|_{\mathbb{R}^n,i} \\ \
= \|\mathbf{W}_o\mathbf{W}_c\|_{\mathbb{R}^n,i} = \|\sqrt{\mathbf{W}_o\mathbf{W}_c}\|^2_{\mathbb{R}^n,i} \\ \
= \sigma_{\max}^2(\sqrt{\mathbf{W}_o\mathbf{W}_c}).
\end{array}
\end{equation}
In fact, if the system is controllable and observable, the entire nonzero spectrum of 
$\Gamma_G^{\dagger}\Gamma_G$ can be obtained.
\begin{equation}
\label{eq:hsv1}
\begin{array}{l}
\textrm{nonzero squared singular values of} \ \ \Gamma_G \ 
= \ \textrm{nonzero eigenvalues of} \ \ \Gamma_G^{\dagger}\Gamma_G \\ \
=  \ \textrm{nonzero eigenvalues of} \ \ \Psi_c^{\dagger}\Psi_o^{\dagger}\Psi_o\Psi_c \
=  \ \textrm{eigenvalues of} \ \ \Psi_o^{\dagger}\Psi_o\Psi_c\Psi_c^{\dagger} \\ \
=  \ \textrm{eigenvalues of} \ \ \mathbf{W}_o\mathbf{W}_c \
=  \ \textrm{eigenvalues of} \ \ \mathbf{W}_c\mathbf{W}_o 
\end{array}
\end{equation}
The singular values of $\Gamma_G$ are called the Hankel singular values (HSV). The nonzero
HSV are the eigenvalues of $\sqrt{\mathbf{W}_c\mathbf{W}_o}$, the set of invariants mentioned
at the end of the previous subsection.

The Hankel norm model reduction problem
is useful for finding bounds for the full $\mathcal{H}_{\infty}$ model reduction problem. 
In particular, it establishes the HSV as a measure of response.
\begin{definition}[Hankel Norm Model Reduction Problem]
Given a rank $n$ Hankel operator $\Gamma_G$ corresponding to a stable, causal, LTI system $G$, 
such that $\Gamma_G:\mathcal{L}^m_2[-T,0] \to \mathcal{L}^p_2[0,T]$,
find $\inf_{\mathrm{rank}(\tilde{\Gamma})\le k} \| \Gamma_G - \tilde{\Gamma}\|_{\mathcal{L}_2,i}$ 
for $\tilde{\Gamma}$
a Hankel operator and $k<n$.
\end{definition}
In order to motivate the solution to the above problem,
we first need to introduce the following theorem.

\begin{theorem}
Given a rank $n$ matrix $\mathbf{M} \in \mathbb{R}^{p\times r}$ $(n \le \min(p,r))$
with nonzero singular values ordered such that 
$\sigma_1(\mathbf{M})\ge\sigma_2(\mathbf{M})\ge \hdots \ge\sigma_n(\mathbf{M})$,
for an arbitrary rank $m$ matrix $\mathbf{S} \in  \mathbb{R}^{p\times r}$ such that 
$m \le k < n$,
\begin{equation} 
\sigma_{\max}(\mathbf{M}-\mathbf{S}) \ge \sigma_{k+1}(\mathbf{M})
\end{equation}
\label{thm:inequal1}
\end{theorem}
Combining \eqref{eq:G_decomp} and \eqref{eq:hankel_norm} and  Theorem \ref{thm:inequal1} leads to 
the next theorem. This is fundamental to this paper, for   
it solves the Hankel model reduction problem.
\begin{theorem}
Given a rank $n$ Hankel operator $\Gamma_G$ 
with nonzero singular values ordered such that 
$\sigma_1(\Gamma_G)\ge\sigma_2(\Gamma_G)\ge \hdots \ge\sigma_n(\Gamma_G)$,
for an arbitrary rank $k$ Hankel operator $\tilde{\Gamma}$ such that  
$k < n$,
\begin{equation} 
\label{eq:hlb2}
\| \Gamma_G-\tilde{\Gamma} \|_{\mathcal{L}_2,i} \ge \sigma_{k+1}(\Gamma_G)
\end{equation}
\end{theorem}
A limitation of this theorem is that, for finite dimensional systems,
there does not always exist a Hankel operator that makes the inequality an equality.

When equation \eqref{eq:hlb2} is combined with $\Gamma_G = P_+G|_{\mathcal{L}_2[-T,0]}$,
we obtain a lower bound for $G$.
\begin{equation}
\label{eq:hinfin_lb}
\begin{array}{c}
\textrm{For}\ \mathrm{order}(G) = n \ \textrm{(i.e. McMillan Degree $n$)}, \\
 \textrm{For all} \ \tilde{G} 
\ \textrm{of order} \ k \le r, 
\| G-\tilde{G} \|_{\mathcal{L}_2,i} \ge \sigma_{r+1}(\Gamma_G)
\end{array}
\end{equation}
An interpretation of this lower bound is that the best 
possible $r^{\textrm{th}}$-order approximation
to the input-output behavior of the system is at least a \lq\lq distance'' $\sigma_{r+1}(\Gamma_G)$
away from the exact response.  
It implies that
any reduced order model that projects out states corresponding to large singular values is necessarily
a worse approximation than a model that projects out small singular values.  Thus, 
states corresponding to large singular values contribute the most to
the system's response.  

$\bullet$ {\bf Balanced realizations}

It has been shown that the nonzero HSV correspond to the eigenvalues of 
 $\sqrt{\mathbf{W}_c\mathbf{W}_o}$.  This suggests that the Hankel operator is 
related to the system's controllability and observability.
This connection is important for many reasons.  Firstly, 
interpreting
the Hankel operator in terms of observability and controllability aids intuition.  Secondly, 
the gramians' eigenvalues are not invariant under coordinate transformations, 
so we still lack unambiguous measures of controllability and observability. 
Lastly, as we can see in Figure \ref{fig:gramians},
controllability and observability may not be correlated.
For generic realizations, observability and controllability are not on the same footing and consequently
this leads to further ambiguity.  Should model reduction be based on observability or controllability?

The resolution to these problems relies on determining the most suitable coordinates.
This is equivalent to
ascertaining the proper way to coarse grain the system. 
Our freedom in the choice of coordinates 
allows us 
to find a coordinate transformation, $\mathbf{T}$, such that, in the new coordinates, 
the controllability and observability gramians are equal and diagonal.
The reader may note that this procedure
is essentially the same as is used in filtering theory.  This 
aligns the observability and controllability ellipsoids, thereby putting controllability
and observability on the same footing.  Furthermore, in these \emph{balanced coordinates},
$\widetilde{\mathbf{W}}_c = \widetilde{\mathbf{W}}_o = \mathbf{\Sigma}$ where
 $\mathbf{\Sigma}$ is diagonal
and has the same eigenvalues as $\sqrt{\mathbf{W}_c\mathbf{W}_o}$ (ordered from largest to smallest).  
The eigenvalues of the balanced gramians are the nonzero HSV, invariants of the system.  
The resulting system realization is known as a \emph{balanced realization}.  

$\mathbf{T}$ can be constructed using the following algorithm \cite{robust1}.
By definition, there exists 
a coordinate transformation $\mathbf{S}$ such that 
$\mathbf{S}^{-1}\mathbf{W}_c\mathbf{W}_o\mathbf{S} = \mathbf{\Sigma}^2$.
Now supposing that $\mathbf{S} = \mathbf{W}_o^{-1/2}\mathbf{R}$ for some $\mathbf{R}$,  
$\mathbf{R}^{-1}\mathbf{W}_o^{1/2}\mathbf{W}_c\mathbf{W}_o^{1/2}\mathbf{R} = \mathbf{\Sigma}^2$. Hence, 
$\mathbf{W}_o^{1/2}\mathbf{W}_c\mathbf{W}_o^{1/2}$ is Hermitian and similar to $\mathbf{\Sigma}^2$.  
Provided $\mathbf{\Sigma}$ does not have degenerate eigenvalues, there exists a unique
unitary matrix $\mathbf{U}$ such that  
$\mathbf{U}^{\dagger}\mathbf{W}_o^{1/2}\mathbf{W}_c\mathbf{W}_o^{1/2}\mathbf{U} = \mathbf{\Sigma}^2$.  
This means that 
\begin{displaymath}
(\mathbf{\Sigma}^{-1/2}\mathbf{U}^{\dagger}\mathbf{W}_o^{1/2})\mathbf{W}_c
(\mathbf{\Sigma}^{-1/2}\mathbf{U}^{\dagger}\mathbf{W}_o^{1/2})^{\dagger} = \mathbf{\Sigma}.
\end{displaymath}  
Thus, remembering that $\mathbf{W}_c$
transforms as  $\mathbf{W}_c \stackrel{\mathbf{T}}{\longrightarrow} 
\widetilde{\mathbf{W}}_c$$=\mathbf{T}^{-1}\mathbf{W}_c(\mathbf{T}^{\dagger})^{-1}$, if we let
$\mathbf{T}^{-1} = \mathbf{\Sigma}^{-1/2}\mathbf{U}^{\dagger}\mathbf{W}_o^{1/2}$, 
we have found the desired coordinate transformation.  It follows that:
\begin{equation}
\label{eq:consistency_check}
\begin{array}{c}
\mathbf{T}^{\dagger}\mathbf{W}_o\mathbf{T} = \big(\mathbf{\Sigma}^{1/2}\mathbf{U}^{\dagger}\mathbf{W}_o^{-1/2}\big)
\mathbf{W}_o\big(\mathbf{W}_o^{-1/2}\mathbf{U}\mathbf{\Sigma}^{1/2}\big) \\
= \mathbf{\Sigma}^{1/2}\mathbf{U}^{\dagger}\mathbf{U}\mathbf{\Sigma}^{1/2} = \mathbf{\Sigma}.
\end{array}
\end{equation}
In general, if a system is not controllable and observable, 
such a $\mathbf{T}$ (balancing transformation) does not exist.
It is possible to find a balancing transformation such that:
\begin{equation}
\label{eq:full_balancing}
\mathbf{W}_c = \left[ \begin{array}{cccc}
\mathbf{\Sigma} & 0 & 0 & 0 \\
0 & \mathbf{\Sigma}_1 & 0 & 0 \\
0 & 0 & 0 & 0 \\
0 & 0 & 0 & 0 
\end{array} \right] \textrm{and} \quad
\mathbf{W}_o = \left[ \begin{array}{cccc}
\mathbf{\Sigma} & 0 & 0 & 0 \\
0 & 0 & 0 & 0 \\
0 & 0 & \mathbf{\Sigma}_2 & 0 \\
0 & 0 & 0 & 0 
\end{array} \right],
\end{equation}
where $\mathbf{\Sigma}, \mathbf{\Sigma}_1,$ and $\mathbf{\Sigma}_2$ are all diagonal.  
$\mathbf{\Sigma}$ is the matrix of HSV
and the corresponding subsystem is controllable and observable.
The subsystem associated with $\mathbf{\Sigma}_1$ is controllable and unobservable, while
the one associated with $\mathbf{\Sigma}_2$ is uncontrollable and observable.

$\bullet$ {\bf Balanced truncation}

Now we possess the tools to generate
reduced-order models.
The reduction technique in what follows is called \emph{balanced truncation}.
We assume that the system is stable, controllable and observable,
and has been transformed into a balanced realization.  Additionally, since the system
is stable,  we consider 
the problem over an infinite time horizon (i.e. $T \to \infty$).

Given a system satisfying the above assumptions, decompose the matrix of ordered HSV $\mathbf{\Sigma}$ 
(ordered from largest to smallest)
such that the first $r$ eigenvalues
form the matrix $\mathbf{\Sigma}_L$. 
The remainder form the matrix of smaller singular values $\mathbf{\Sigma}_S$.  Decompose
 $\mathbf{\Sigma}_L$ and $\mathbf{\Sigma}_S$ such that they have no common eigenvalues. The realization
for the full system takes the form:
\begin{equation}
\label{eq:bal_real}
\begin{array}{c}
\tilde{\mathbf{A}} = \left[\begin{array}{cc} 
\tilde{\mathbf{A}}_L & \tilde{\mathbf{A}}_{12} \\ 
\tilde{\mathbf{A}}_{21} & \tilde{\mathbf{A}}_S
\end{array}\right], \ \  
\tilde{\mathbf{B}} = \left[\begin{array}{c}
\tilde{\mathbf{B}}_L \\ \tilde{\mathbf{B}}_S
\end{array}\right],
\\
\tilde{\mathbf{C}} = \left[\begin{array}{cc}
\tilde{\mathbf{C}}_L & \tilde{\mathbf{C}}_S
\end{array}\right].
\end{array}
\end{equation}
By projecting out the states corresponding to $\mathbf{\Sigma}_S$, 
 the remaining three matrices,
$(\tilde{\mathbf{A}}_L,\tilde{\mathbf{B}}_L,\tilde{\mathbf{C}}_L)$,
 form an $r$-dimensional realization that approximates the original system.  Denote the input-output
operator for the reduced system by $\tilde{G}_r$. 
This realization, by construction, is stable and balanced.
Its HSV are the eigenvalues of $\mathbf{\Sigma}_L$. 
In additional,
the approximation error is given by:
\begin{equation}
\label{eq:up_bound}
\| G - \tilde{G}_r \|_{\mathcal{L}_2,i} \le 2\sum_{j=1}^k \sigma^{\mathrm{dist}}_{i_j},
\end{equation}
where $\{\sigma^{\mathrm{dist}}_{i_j}: 1 \le j \le k \}$ is the distinct HSV in $\mathbf{\Sigma}_S$.

These techniques are extended to unstable systems in Appendix \ref{sec:infiniteH}.
This makes it possible to use these techniques on linear,
conservative systems.  A particularly relevant result is:
\begin{theorem}[Lower Bound]
Given a LTI, causal system $G$ with $n$ dimensional minimal realization $(\mathbf{A},\mathbf{B},\mathbf{C})$.
If there exists an \lq\lq$a$'' such that $-a\mathbf{I}+\mathbf{A}$ is a stable system matrix then 
for any order $r$ (or less) approximant $\tilde{G}_r$
\begin{displaymath} 
\displaystyle \| G-\tilde{G}_r \|_{\mathcal{L}_2[0,T],i} \ge 
\big(1-e^{-2aT}\|e^{\mathbf{A}^{\dagger}T}e^{\mathbf{A}T}\|_{\mathbb{C}^n,i}\big)\sigma_{r+1}(a)
\end{displaymath}
\end{theorem}
It is the subject of the next section
to apply these methods to general oscillator systems, 
whereupon, when combined with the spatial content of the reductions, 
specifies how to coarse grain.


\section{Reduction of Oscillator Systems}
\label{sec:Ham}

The standard form for the equations of motion generated by a quadratic Hamiltonian 
with $2N$ phase space degrees of freedom is given by
\begin{equation}
\label{eq:lin_ham}
\left[ \begin{array}{c}
\dot{q} \\ \dot{p}  
\end{array} \right] =
\left[ \begin{array}{cc} 
0 &\mathbf{\Omega} \\  - \mathbf{\Omega} & 0 
\end{array} \right]
 \left[ \begin{array}{c}
q \\ p  
\end{array} \right],
\end{equation}
where $\mathbf{\Omega}$ is a $N \times N$ positive definite matrix.  These systems are typically associated with 
coupled harmonic oscillators. Furthermore, these systems are  considered trivial
because when expressed in normal modes, 
 the resulting oscillators are decoupled.
Decoupled oscillators are considered noninteracting.
This view is correct for isolated systems, however, for open systems it is not.  

The coordinates that capture the original experimental 
configuration is of exceptional interest.
This is because 
the important coordinate system for model reduction 
is the one in which the gramians are balanced.  
Now while in balanced coordinates
the gramians are diagonalized, the matrices $\mathbf{A}$ or  $\mathbf{\Omega}$
need not be.  This illustrates that a generic open system, even a linear one,
typically has interacting internal states.
The statistical mechanics of quadratic Hamiltonians is invalid unless the 
system is driven.  This follows since when expressed in normal modes, the system is noninteracting 
and, hence, not ergodic or mixing \cite{vankampen,kubo}.
The usual heuristic argument for justifying the practice
is that phonon (oscillator) systems do not truly have a  linear 
dispersion; the systems themselves are nonlinear. 
The nonlinearity is responsible for the mixing of states.
This is precisely the issue that spawned the Fermi-Pasta-Ulam (FPU) problem 
\footnote{It is interesting to note that coupled, nonlinear, anharmonic oscillator systems are not guaranteed to mix. This fact has been attributed to the existence soliton solutions to the equations of motions.}.   
Once the nonlinearity from the heuristic argument is associated to a disturbance of the form $\mathbf{B}u(t)$,
 then the analysis in this paper agrees with the heuristic argument.  

The advantages of using balanced coordinates rather than modal coordinates reflect the sensitivity
of the system to the choice of experiment.
This sensitivity to experiment suggests that it is  more appropriate for oscillator 
systems \footnote{At least oscillator systems with uniform masses.} to investigate open systems. 
\begin{equation}
\label{eq:gen_ham}
\begin{array}{c}
\dot{z} = 
\left[ \begin{array}{c}
\dot{z}_1 \\ \dot{z}_2  
\end{array} \right] = \mathbf{A}z + \mathbf{B}u = 
\left[ \begin{array}{cc} 
0 &\mathbf{I} \\  - \mathbf{\Omega}^2 & 0 
\end{array} \right]
 \left[ \begin{array}{c}
z_1 \\ z_2  
\end{array} \right]
+ \mathbf{B}u \\
y = \mathbf{C}z
\end{array}
\end{equation}
In this coordinate system, 
 $z_1$ represents the spring displacements 
while $z_2$ represents the corresponding velocities.

There remains the question of which experiments should be considered. 
Conceptually, $\mathbf{B}$ 
varies the gains of the driving. 
$\mathbf{B}$ determines how accessible particular states are to driving.  Alternatively,
allowing for Dirac delta driving  and setting 
$z_0 = 0$, the driving may be used to prepare the system's 
initial conditions.  With this interpretation, $\mathbf{B}$ prescribes how 
initial conditions are weighted.  Similarly, $\mathbf{C}$ indicates which and how easily
internal states are measured.
We exclusively consider oscillator systems with uniform constituent sizes and masses.
This choice, motivated by thermodynamics and the equipartition of energy, 
implies that positions and momenta are treated equally.
This is equivalent to  assuming that the position and momentum of each particle
can be driven with equal gain.  Hence, $\mathbf{B}=b\mathbf{I}$, where $b$ is a constant.  
Considering each
position and momentum equally difficult to measure corresponds to setting $\mathbf{C} = c\mathbf{I}$.
Now let $b=c=1$.  Mathematically this choice is natural since 
the resulting input-output operator (in the Laplace domain) is the full system's resolvent,
$(s\mathbf{I}-\mathbf{A})^{-1}$.  

\begin{figure}
\centerline{\epsfxsize=6cm \epsfbox{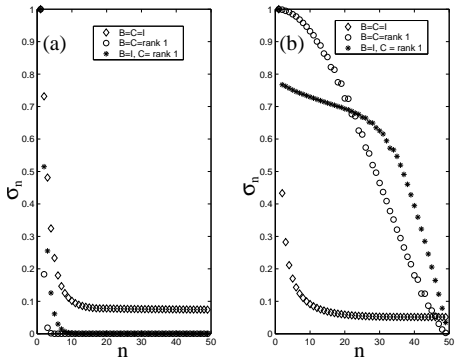}}
\caption{  { \small A plot of the ordered Hankel singular values (HSV) for the homogeneous, harmonic oscillator
chain.  $T$ is the time over which the system is investigated. The HSV in (a) are plotted 
for $T \propto N^{1/2}$, while $T \propto N^{2}$ for (b), where $N$ is the total number of masses in 
the chain. In each case $N = 49$.}}
\label{fig:HSV1}
\end{figure}

Fixing $\mathbf{B}$ and $\mathbf{C}$ dictates the type of experiment.  However, it 
does not fully determine the experiment.  For the problem to be well-posed, we also 
need to specify the duration of the experiment.  
This is important because the exact form of the experiment fully determines 
its input-output characteristics (i.e. the input-output operator).  Different 
experiments give rise to different measures of response. This is illustrated 
in Figure \ref{fig:HSV1}.  As a reminder, Hankel singular values (HSV) tell us
how much their corresponding  states contribute to the response.  These states 
are roughly the system's input-output resonances.

Figure \ref{fig:HSV1}(b) 
depicts how different types of experiments on an oscillator system, over the same 
duration, $T = N^2$, give rise to different normalized HSV and, consequently, different measures of response.  
Figure \ref{fig:HSV1}(a) displays the same types of experiments, but with  $T = N^{1/2}$.
Clearly, the normalized HSV for two of the experiments have completely changed.
Figure \ref{fig:HSV1}(a) demonstrates that  experiments over the shorter time
frames tend to admit lower-order reduced models.
Hence, time scales 
have an enormous impact on model reduction.  

Although experiments over short time horizons directly lead to very low-order reduced models
\cite{barahona02}, our intent for introducing a finite  
cutoff time is to regulate divergences.  The divergences arise due to the fact that the HSV become 
infinite in the infinite-time horizon limit. This is discussed in more detail in Appendix \ref{sec:infiniteH}.
Now with the divergences regulated, we  return to 
addressing how to coarse grain physical systems.  From above, we learn that 
this is not a well-posed question since 
different time scales  or
experiments 
lead to different reductions.  A well-posed reduction
problem requires  specifying the type of experiment and the time scale.
It is natural to expect that past a certain 
time scale (i.e. past thermodynamic equilibration), 
there is a unique way 
to coarse grain.
It is for this reason that we approximate the response of the system in equation
\eqref{eq:gen_ham} with $\mathbf{B}=\mathbf{C} = \mathbf{I}$ over a finite yet long time horizon.
Interestingly, not only does there exist such a time scale, but 
some Hankel norm results
determine precisely the coarse grainings.  These topics 
comprise the following subsections, the first in which we work without 
restriction on the form of $\mathbf{\Omega}$
other than that it is positive definite.  In the second subsection we consider the case  
of the homogeneous linear harmonic chain, and lastly we treat some
heterogeneous linear oscillator chains. 

\subsection{General Oscillator Systems}
\label{subsec:gen_osc}

To determine the best possible coarse graining or at least 
near the optimal coarse graining, we will proceed to use the Hankel operator 
machinery to obtain bounds 
for $\| G  - \tilde{G}_r\|_{\mathcal{L}_2[0,T],i}$.  Recalling the control theory
tutorial, this provides us with a criterion for model reduction.
In particular, the lower bound,
\begin{displaymath}
\| G-\tilde{G} \|_{\mathcal{L}_2,i} \ge \sigma_{r+1}(\Gamma_G),
\end{displaymath}
where $\sigma_{r+1}(\Gamma_G)$ is the $(r+1)^{\mathrm{th}}$ HSV, 
confirms that the states with large HSV contribute the most to the response.
The upper bound,
\begin{displaymath}
\| G-\tilde{G} \|_{\mathcal{L}_2,i} \le 2\sum_{j=1}^k \sigma^{\mathrm{dist}}_{i_j},
\end{displaymath}
where $\{\sigma^{\mathrm{dist}}_{i_j}: 1 \le j \le k \}$ is the set of distinct HSV with $i_j > r$,
ensures that our approximations are controlled.
The first step is to determine the HSV that provide these bounds. 
However, in doing so we also find the balancing transformation.
This makes it trivial to truncate the system and obtain reduced-order models.

Determining the HSV requires calculating
the controllability and observability gramians.
Unfortunately, restricting attention to a finite cutoff time 
 complicates the analysis.  For instance, $\mathbf{W}_c$ and $\mathbf{W}_o$ satisfy 
differential equations (see equations \eqref{eq:diff_lyap1} and \eqref{eq:diff_lyap2}) 
instead of Lyapunov equations.  
A method of simplifying the analysis
involves investigating 
the system matrix $-a\mathbf{I}+\mathbf{A}$ over an infinite time horizon
instead of $\mathbf{A}$ over a finite time horizon.  This procedure 
is known as exponential discounting.
Intuitively \lq\lq$a$'' should be on the order of the inverse time cutoff
for the approximation to be any good.  Fortunately the above intuition can be made much 
more rigorous, 
thereby keeping all approximations under control.  

For the systems under consideration, the gramians are formally given by
\begin{equation}
\label{eq:grams}
\begin{array}{c}
\mathbf{W}_c = \int_0^T e^{\mathbf{A}t}e^{\mathbf{A}^{\dagger}t} dt \ \Rightarrow
\mathbf{W}_c^{(a)} = \int_0^{\infty} e^{-2at}e^{\mathbf{A}t}e^{\mathbf{A}^{\dagger}t} dt, \\
\mathbf{W}_o = \int_0^T e^{\mathbf{A}^{\dagger}t}e^{\mathbf{A}t} dt \ \Rightarrow
\mathbf{W}_o^{(a)} = \int_0^{\infty} e^{-2at}e^{\mathbf{A}^{\dagger}t}e^{\mathbf{A}t} dt.
\end{array}
\end{equation}
A property of these gramians is that
$\mathbf{W}_c\mathbf{W}_o$ or $\mathbf{W}_c^{(a)}\mathbf{W}_o^{(a)}$ are always 
similar to a matrix of the form 
$\left[\begin{array}{cc} \mathbf{M} & 0 \\ 0 & \mathbf{M}^{\dagger} \end{array}\right]$.
This means that there is an exact duplicity in the HSV independent of \lq\lq$a$''. 
In fact, under the transformation $\mathbf{R}$
defined in Appendix \ref{sec:calculation}, $\mathbf{A}$  takes  the 
form of the system matrix in equation \eqref{eq:lin_ham}.
In this basis, we find that $\widetilde{\mathbf{W}}_c \approx \widetilde{\mathbf{W}}_o$ where 
\begin{equation}
\label{eq:pre_result}
\begin{array}{c}
\widetilde{\mathbf{W}}_c = \frac{1}{4a}\left[\begin{array}{cc}
\mathbf{\Omega} + \mathbf{\Omega}^{-1} + \mathcal{O}(a^2) & 0 \\ 
0 & \mathbf{\Omega} + \mathbf{\Omega}^{-1} + \mathcal{O}(a^2) \end{array}\right] \\
+ \frac{1}{4}\left[\begin{array}{cc}
0 & \mathbf{\Omega}^{-2}-\mathbf{I} + \mathcal{O}(a^2) \\
 \mathbf{\Omega}^{-2}-\mathbf{I} + \mathcal{O}(a^2) & 0 \end{array}\right].
\end{array}
\end{equation}
In this basis the gramians are almost balanced.  Provided we transform the system by a unitary 
transformation, $\mathbf{U}_d$, to diagonalize $\mathbf{\Omega}$ (i.e. 
$\mathbf{\Omega}\stackrel{\mathbf{U}_d}{\longrightarrow} \mathbf{\Lambda}_{\Omega}$)
and we take \lq\lq$a$'' sufficiently small (sufficiently long-time horizons),
 the gramians are balanced.  
The precise interpretation of 
``sufficiently small'' is outlined in Appendix \ref{sec:restrictions}. 
We want to require that \lq\lq$a$'' is small enough so that the off-diagonal terms do not 
change the ordering of the HSV. 
Here we assume, without loss of generality, that the eigenvalues of $ \mathbf{\Lambda}_{\Omega}$ 
are ordered from smallest to largest. 
Under the previously mentioned conditions to $\mathcal{O}(a^2)$
the balanced gramians (balanced up to permutation) take the form
\begin{equation}
\label{eq:main_result}
\bar{\mathbf{W}}_c = \bar{\mathbf{W}}_o = 
 \frac{1}{4a}\left[\begin{array}{cc}
\mathbf{\Lambda}_{\Omega} + \mathbf{\Lambda}_{\Omega}^{-1} + \mathcal{O}(a^2) & \mathcal{O}(a) \\ 
\mathcal{O}(a) & \mathbf{\Lambda}_{\Omega} + \mathbf{\Lambda}_{\Omega}^{-1} + \mathcal{O}(a^2)  \end{array}\right].
\end{equation}
These balanced gramians are associated with  the linear system
\begin{equation}
\label{eq:bal_real_main}
\begin{array}{c} 
\left[ \begin{array}{c}
\dot{\bar{z}}_1 \\ \dot{\bar{z}}_2  
\end{array} \right] =
\left[ \begin{array}{cc} 
0 &\mathbf{\Lambda}_{\Omega} \\  - \mathbf{\Lambda}_{\Omega} & 0 
\end{array} \right]
 \left[ \begin{array}{c}
\bar{z}_1 \\ \bar{z}_2  
\end{array} \right]
+ \left[ \begin{array}{cc} \mathbf{\Lambda}_{\Omega}^{1/2}\mathbf{U}_d^{\dagger} & 0 \\
0 & \mathbf{\Lambda}_{\Omega}^{-1/2}\mathbf{U}_d^{\dagger} \end{array} \right] 
\left[ \begin{array}{c} u_1 \\ u_2 \end{array} \right], \\
\left[ \begin{array}{c} y_1 \\ y_2 \end{array} \right] = 
\left[ \begin{array}{cc} \mathbf{U}_d\mathbf{\Lambda}_{\Omega}^{-1/2} & 0 \\
0 & \mathbf{U}_d\mathbf{\Lambda}_{\Omega}^{1/2} \end{array} \right] 
\left[ \begin{array}{c} \bar{z}_1 \\ \bar{z}_2 \end{array}\right].
\end{array}
\end{equation}
Given that $\lambda_j(\mathbf{\Omega}) \le \lambda_k(\mathbf{\Omega})$ for all $j<k$, let 
 $\alpha$ be a permutation such that 
 $\lambda_{\alpha(j)}(\mathbf{\Omega})+\lambda_{\alpha(j)}^{-1}(\mathbf{\Omega}) \le 
\lambda_{\alpha(k)}(\mathbf{\Omega})+\lambda_{\alpha(k)}^{-1}(\mathbf{\Omega})$ for all $k<j$.
Trivially, via a unitary transformation,
the gramians in equation \eqref{eq:main_result} may be fully balanced
(the HSV are ordered).  In this case, we find that
\begin{equation}
\label{eq:HSV_ordering}
\sigma_{k}(a) = \frac{1}{4a}\big(\lambda_{\lceil \frac{\alpha(k)}{2} \rceil}(\mathbf{\Omega}) + 
\lambda_{\lceil \frac{\alpha(k)}{2} \rceil}^{-1}(\mathbf{\Omega})\big) + \mathcal{O}(1).
\end{equation}

The degeneracy of the HSV suggests that any balanced truncation that keeps states corresponding to 
the first $r=2q$ HSV will remain conservative.  In fact, we can see this by inspection of 
equation \eqref{eq:bal_real_main}. 
With this in mind, 
we will 
consider only truncations that keep an even number of states.  
The immediate consequences 
of these results are that 
we obtain bounds on the approximation error of the response.
\begin{equation}
\label{eq:gm_bounds1}
\begin{array}{c}
\displaystyle \| G  - \tilde{G}_{2q}\|_{\mathcal{L}_2[0,T],i} \ge \big(1-e^{-2aT}\big)\sigma_{2q+1}(a) \\
\displaystyle = \frac{1}{4a}(1-e^{-2aT})\big( \lambda_{\alpha(q+1)}(\mathbf{\Omega})+\lambda_{\alpha(q+1)}^{-1}(\mathbf{\Omega}) \big),
\end{array}
\end{equation}
and
\begin{equation}
\label{eq:gm_bounds2}
\begin{array}{c}
\displaystyle \| G  - \tilde{G}_{2q}\|_{\mathcal{L}_2[0,T],i} \le 2e^{aT}\sum_{j=q}^{N-1} \sigma_{2j+1}(a) \\
\displaystyle = \frac{1}{2a}e^{aT}\sum_{j=q+1}^{N} \lambda_{\alpha(j)}(\mathbf{\Omega})+\lambda_{\alpha(j)}^{-1}(\mathbf{\Omega}).
\end{array}
\end{equation}
Additionally, by using linear-matrix-inequality (LMI) techniques \cite{boyd}, 
a substantially tighter upper bound can be established.
The improved bound is
\begin{equation}
\label{eq:main_upbound}
 \begin{array}{c}
\| G  - \tilde{G}_{2q}\|_{\mathcal{L}_2[0,T],i} \le 2e^{aT}\sigma_{2k+1}(a) \\
= \frac{1}{2a}e^{aT}\big(\lambda_{\alpha(k+1)}(\mathbf{\Omega})+\lambda_{\alpha(k+1)}^{-1}(\mathbf{\Omega})\big).
\end{array}
\end{equation}

A remarkable aspect of these oscillator models is that, over a long time horizon, 
the relation between the system's spectrum and the gramians is simple.  Despite this 
simple relationship, 
these results also establish that 
the set of best reductions (i.e. those that satisfy the lower bound) need not be modal reductions!
Modal reductions explicitly neglect (project out) the system's fast dynamics.
For instance, let us suppose that $\lambda_{j}(\mathbf{\Omega})\ge 1$ for all $j$; 
disregarding degeneracy, the HSV are automatically ordered from smallest to largest.  This implies
that projecting out small HSV eliminates states  corresponding to slow modes!  
This is contrary to what is typically done.
Alternatively, suppose that  $\lambda_{j}(\mathbf{\Omega}) < 1$ for all $j$.  In this instance, disregarding
degeneracy, the HSV are ordered from largest to smallest.  This is precisely 
when modal reduction is appropriate.  
Lastly, when the eigenvalues of $\mathbf{\Omega}$ are both greater than and less than one,
the appropriate reductions
involve a mixing of fast and slow modes.

In the basis that produces the realization in 
equation \eqref{eq:bal_real_main}, the system is balanced, and yet $\mathbf{\Omega}$
is diagonalized.  
Although this means that
such systems are noninteracting,  
not all internal states are treated equally.  
In fact, in the case of the linear harmonic chain, the weighting of the gains has a 
physically meaningful interpretation that will be elaborated upon in Section \ref{subsec:lin_chain}.  
Also, there is an
enormous degeneracy in the types of experiments 
producing equivalent reductions.
The same reductions result from using
$\mathbf{B} = \mathbf{V}$ and $\mathbf{C} = \mathbf{U}$ 
where $\mathbf{V}$ and $\mathbf{U}$ are arbitrary unitary matrices.
This may not come as much of a surprise 
since requiring $\mathbf{V}$ and $\mathbf{U}$ to be unitary causes the internal
states to be treated equally.  
We see that there are an infinite number of inequivalent realizations
that yield the same reduction. 

These results also reveal 
how to choose \lq\lq$a$''.  By varying \lq\lq$a$'' we may refine our bounds.  
Generally, we have an
LTI, causal system 
that achieves or almost achieves the lower bound.  Consequently, it is useful to choose \lq\lq$a$''
such that the lower bound is at its maximum. 
The maximum occurs as 
$a \to 0$.
 \begin{equation}
\label{eq:last_lb}
\| G  - \tilde{G}_{2q}\|_{\mathcal{L}_2[0,T],i} \ge 
 \frac{T}{2}\big(\lambda_{\alpha(q+1)}(\mathbf{\Omega})+\lambda_{\alpha(q+1)}^{-1}(\mathbf{\Omega})\big)
\end{equation}
For the lowest upper bound (from balanced truncation),
$\frac{\mathrm{d}}{\mathrm{d}a}\big[ 2e^{aT}\sigma_{2k+1}(a)\big] = 0$ implies that $a=T^{-1}$.
Therefore, the minimum upper bound is
 \begin{equation}
\label{eq:last_ub}
\| G  - \tilde{G}_{2q}\|_{\mathcal{L}_2[0,T],i} \le 
 \frac{eT}{2}\big(\lambda_{\alpha(q+1)}(\mathbf{\Omega})+\lambda_{\alpha(q+1)}^{-1}(\mathbf{\Omega})\big).
\end{equation}

As we approximate $G$ over an infinite-time horizon ($T \to \infty$),
both the norm of $G$ and the absolute error diverge. 
This divergence is unavoidable and generally independent of the number of particles (oscillators)
in the system.  However, there is another possible interpretation if  the number of
particles, $N$, is large.
For our analytics to be valid, there must be restrictions on the size of \lq\lq$a$''.  
Combining equations \eqref{eq:pert8} and \eqref{eq:a_restriction} from
Appendix \ref{sec:restrictions}, we are able to relate 
the maximal \lq\lq$a$'' 
to the frequencies of the system, $\lambda_{k}(\mathbf{\Omega})$.
In most cases, as $N$ gets large, $\lambda_{k}(\mathbf{\Omega}) \sim  N^{-\delta_1}$.
For instance, in the case of the homogeneous linear harmonic chain, for small wave number,
$\delta_1 = 1$.  Consequently, \lq\lq$a$'' is forced to fall to zero asymptotically like
$a \sim N^{-\delta_2}$ for some $\delta_2 >0$ as $N \to \infty$.
Reciprocally, if \lq\lq$a$'' tends to zero faster and, consequently, $T$ tends to infinity, there is no 
restriction on $N$. The divergence is due to the infinite time horizon.
Suppose that 
\lq\lq$a$'' is parameterized by $N$ and we consider 
the $N \to \infty$ limit.  In this case, the divergence is due
to the infinite-particle limit.  For oscillator systems, like photons
and phonons, the divergence is attributable to the absence of a mass gap
(i.e. the eigenvalues of $\mathbf{\Omega}$ become dense near zero).  
Thus, there is no inherent length (mass) scale for the system.  This is 
one of the simplest 
divergences, a long wavelength divergence.
Depending on the structure of $\mathbf{\Lambda}_{\Omega}$, however,
there may also be a short-wavelength divergence 
or even possibly a mixed-wavelength divergence.
Had we investigated yet a shorter time scale, still  taking 
$T \to \infty$, the resulting reduced-order systems 
are typically dissipative \cite{reyn1,barahona02}.
Physically this is a manifestation of how
fluctuations may induce time scales \cite{vankampen,kubo}.

While the divergence in $\| G \|_{\mathcal{L}_2,i}$ and 
 $\| G -\tilde{G}_r\|_{\mathcal{L}_2,i}$ has been explained, its consequences have not.
Since the error estimate diverges, except when regulated
(i.e. by considering finite $T$),
the absolute error is not a meaningful quantity.
Any long-time approximate is asymptotic at best.  This means that the (regulated) relative error,
 $\lim_{T \to \infty}\| G -\tilde{G}_r\|_{\mathcal{L}_2[0,T],i}/\| G \|_{\mathcal{L}_2[0,T],i}$,
 is much more useful.
Combining equations \eqref{eq:gm_bounds1} and \eqref{eq:gm_bounds2} yield rather conservative bounds:
\begin{equation}
\label{eq:gen_ugly1}
\begin{array}{c}
 \displaystyle \lim_{T \to \infty} \frac{\| G -\tilde{G}_{2q}\|_{\mathcal{L}_2[0,T],i}}{\| G \|_{\mathcal{L}_2[0,T],i}}
\ge \lim_{T \to \infty} \frac{ \frac{T}{2}\big(\lambda_{\alpha(q+1)}(\mathbf{\Omega})+
\lambda_{\alpha(q+1)}^{-1}(\mathbf{\Omega})\big) }{ \frac{eT}{2}
\sum_{j=1}^{N} \lambda_{\alpha(j)}(\mathbf{\Omega})+\lambda_{\alpha(j)}^{-1}(\mathbf{\Omega}) } \\
\displaystyle  =  \frac{ \lambda_{\alpha(q+1)}(\mathbf{\Omega})+\lambda_{\alpha(q+1)}^{-1}(\mathbf{\Omega}) }
{ e \mathrm{Tr}\big( \mathbf{\Omega} + \mathbf{\Omega}^{-1} \big) },
\end{array}
\end{equation}
and
\begin{equation}
\label{eq:gen_ugly2}
\begin{array}{c}
\displaystyle \lim_{T \to \infty} \frac{\| G -\tilde{G}_{2q}\|_{\mathcal{L}_2[0,T],i}}{\| G \|_{\mathcal{L}_2[0,T],i}}
\le \lim_{T \to \infty} \frac{ \frac{eT}{2} \sum_{j=q+1}^{N} \lambda_{\alpha(j)}(\mathbf{\Omega})+
\lambda_{\alpha(j)}^{-1}(\mathbf{\Omega}) }{ \frac{T}{2}\big(\lambda_{\alpha(1)}(\mathbf{\Omega})+
\lambda_{\alpha(1)}^{-1}(\mathbf{\Omega})\big) } \\
\displaystyle =  \frac{ e \sum_{j=q+1}^{N} \lambda_{\alpha(j)}(\mathbf{\Omega})+
\lambda_{\alpha(j)}^{-1}(\mathbf{\Omega}) }
{ \big(\lambda_{\alpha(1)}(\mathbf{\Omega})+\lambda_{\alpha(1)}^{-1}(\mathbf{\Omega})\big) }.
\end{array}
\end{equation}
Comparatively tighter bounds are obtained by using equations \eqref{eq:last_lb} and \eqref{eq:last_ub}.
These bounds are
\begin{equation}
\label{eq:relb1}
\lim_{T \to \infty} \frac{\| G -\tilde{G}_{2q}\|_{\mathcal{L}_2[0,T],i}}{\| G \|_{\mathcal{L}_2[0,T],i}}
\ge  \frac{ \lambda_{\alpha(q+1)}(\mathbf{\Omega})+ \lambda_{\alpha(q+1)}^{-1}(\mathbf{\Omega}) }
{ e \big(\lambda_{\alpha(1)}(\mathbf{\Omega})+\lambda_{\alpha(1)}^{-1}(\mathbf{\Omega})\big) },
\end{equation}
and
\begin{equation}
\label{eq:relb2}
\lim_{T \to \infty} \frac{\| G -\tilde{G}_{2q}\|_{\mathcal{L}_2[0,T],i}}{\| G \|_{\mathcal{L}_2[0,T],i}}
\le  \frac{ e\big(\lambda_{\alpha(q+1)}(\mathbf{\Omega})+ \lambda_{\alpha(q+1)}^{-1}(\mathbf{\Omega})\big) }
{ \lambda_{\alpha(1)}(\mathbf{\Omega})+\lambda_{\alpha(1)}^{-1}(\mathbf{\Omega}) }.
\end{equation}

These general cases have
allowed us, for instance, to determine conditions when 
modal reduction is appropriate.  Without knowing more about  $\mathbf{\Omega}$
it is not possible to discern the spatial content of the reductions.
Without the spatial content we cannot specify the relationship between reduction type 
and coarse graining.  In the following section,
we will apply the above results to the linear harmonic chain,
from which we determine the appropriate coarse grainings.  
This example will clarify
 the relationship between system reductions and coarse grainings.

\subsection{The Linear Harmonic Chain}
\label{subsec:lin_chain}

\begin{figure}
\centerline{\epsfxsize=6cm \epsfbox{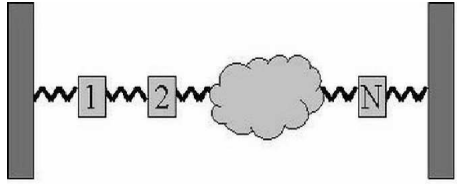}}
\caption{ { \small Linear chain of oscillators with fixed boundary conditions.  All
of our  models are of this form.  The different variations
have homogeneous, layered, and randomly, uniformly sampled spring constants.}}
\label{fig:springs}
\end{figure}

The models that we consider in this section are all variants of the one-dimensional 
linear harmonic chain depicted in Figure \ref{fig:springs}.  The system consists of a chain
of $N$ equally spaced masses each with mass $m$ connected via $N+1$ springs.
The chain is connected on each side to stationary walls.  
We will first treat coarse graining the 
linear harmonic chain with homogeneous springs in great detail.   
Briefly, we also present how to coarse grain some heterogeneous chains.
We will conclude by comparing the different models and their respective coarse grainings.

$\bullet$ {\bf The homogeneous chain}
 
Each spring of the homogeneous linear chain has a spring constant, $k$.
 For this system, 
$\mathbf{\Omega}^2$ has the familiar form,
\begin{equation}
\label{eq:lin_osc_matrix}
\mathbf{\Omega}^2 = \frac{k}{m}
\left[\begin{array}{ccccc}
2 & -1 & 0 & \hdots & 0 \\
-1 & \ddots & \ddots & \ddots & \vdots \\
0 & \ddots & \ddots & \ddots  & 0 \\
\vdots & \ddots & \ddots & \ddots & -1 \\
0 & \hdots & 0 & -1 & 2 
\end{array}\right]
\end{equation}
The matrix of ordered eigenvalues of $\mathbf{\Omega}$,
 $\mathbf{\Lambda}_{\Omega}$, is such that 
$\displaystyle (\mathbf{\Lambda}_{\Omega})_{pp} =\omega_p= 2\sqrt{\frac{k}{m}}\sin\big(\frac{\pi p}{2(N+1)}\big)$.
Additionally, the unitary matrix that diagonalizes $\mathbf{\Omega}$, $\mathbf{U}_d$, is given by 
$\displaystyle (\mathbf{U}_d)_{ij} = (\mathbf{u}^{(j)})_i = \sqrt{\frac{2}{N+1}}\sin\big(\frac{\pi ij}{N+1}\big)$.
Here $\mathbf{u}^{(j)}$ is the eigenvector such that $\mathbf{\Omega}\mathbf{u}^{(j)} = \omega_j\mathbf{u}^{(j)}$.
Not only is  $\mathbf{U}_d$ both orthogonal and symmetric;  its action on vectors
is almost that of a discrete Fourier transform.  It is not actually a Fourier transform since the spatial domain of 
lattice sites is not translationally invariant. Had we considered the linear harmonic chain on 
a ring instead (i.e. the group $\mathbb{Z}_N$), then the action of $\mathbf{U}_d$ on vectors would, in fact, be a 
Fourier transform.  The main point here is that local spatial rescaling in real space corresponds to 
rescaling large wave vectors in Fourier space.

Motivated by model reduction, we consider two particularly interesting limits.
For the first case, let
$\displaystyle 2\sqrt{\frac{k}{m}} \le 1$ and $N \gg 1$.
In the second case we take
the mass and spring constants to be functions of $N$ such that 
$\displaystyle 2\sqrt{\frac{k(N)}{m(N)}} \ge \frac{2(N+1)}{\pi}$ and $N \gg 1$.  
The former case will be discussed
in detail. After that the nuances of latter case will be clear.

In the first case, $\omega_p<1$ for all $p \in \{1,\hdots,N\}$. Consequently, 
$\alpha(p) = p$.  Hence, the HSV are already ordered from largest to smallest.  Also, 
the minimal time scale over which this  
analysis is valid is determined by the limits on \lq\lq$a$''.
When \lq\lq$a$'' satisfies the inequality in equation \eqref{eq:a_restriction}, which implies
\lq\lq$a$'' at least scales as $N^{-2}$ (if not faster in $N$), the ordering of the HSV is not altered.   
This gives the absolute error bounds when combined with the expression for $\omega_p$ and 
 equations \eqref{eq:last_lb} and \eqref{eq:last_ub}.
 \begin{eqnarray}
\label{eq:ho_lb1}
\begin{array}{c}
\| G  - \tilde{G}_{2q}\|_{\mathcal{L}_2[0,T],i} \ge 
 \frac{T}{2}\bigg( \sin\big(\frac{\pi(q+1)}{2(N+1)}\big)+ \Big(\sin\big(\frac{\pi(q+1)}{2(N+1)}\big)\Big)^{-1} \bigg) \\
= \frac{NT}{\pi(q+1)}\Big(1 + \mathcal{O}\big(\left(\frac{q+1}{N}\right)^2\big)\Big)
\end{array}
\\
\label{eq:ho_ub1}
\begin{array}{c}
\| G  - \tilde{G}_{2q}\|_{\mathcal{L}_2[0,T],i} \le 
 \frac{eT}{2}\bigg( \sin\big(\frac{\pi(q+1)}{2(N+1)}\big)+ \Big(\sin\big(\frac{\pi(q+1)}{2(N+1)}\big)\Big)^{-1} \bigg) \\
= \frac{eNT}{\pi(q+1)}\Big(1 + \mathcal{O}\big(\left(\frac{q+1}{N}\right)^2\big)\Big)
\end{array}
\end{eqnarray}

It is no surprise that the appropriate reductions project out the fast modes
since in this limit the dispersion is linear.
In the limit of large $N$, truncating fast modes is the same as projecting out large wave numbers. 
However, as mentioned earlier, large wave numbers 
correspond to short distances.  
So we see that projecting
out fast modes from this system is equivalent to locally coarse graining.
In fact, the lower bound suggests a stronger result.  
Provided the lower bound is approximately
achievable, though the reduced-order system may not be LTI, the \emph{best possible} reductions
must involve locally coarse graining (modally reducing) the system.  
For example,
projecting out a slow mode via balanced truncation is an example of nonlocal coarse graining.
The lower bound of the incurred approximation error involves the HSV
corresponding to that nonlocal state.  Since the HSV corresponding to slow (nonlocal) modes 
are larger than those of fast modes, 
any nonlocal approximant of the system is necessarily worse by equation \eqref{eq:ho_lb1}.

\begin{figure}
\centerline{\epsfxsize=6cm \epsfbox{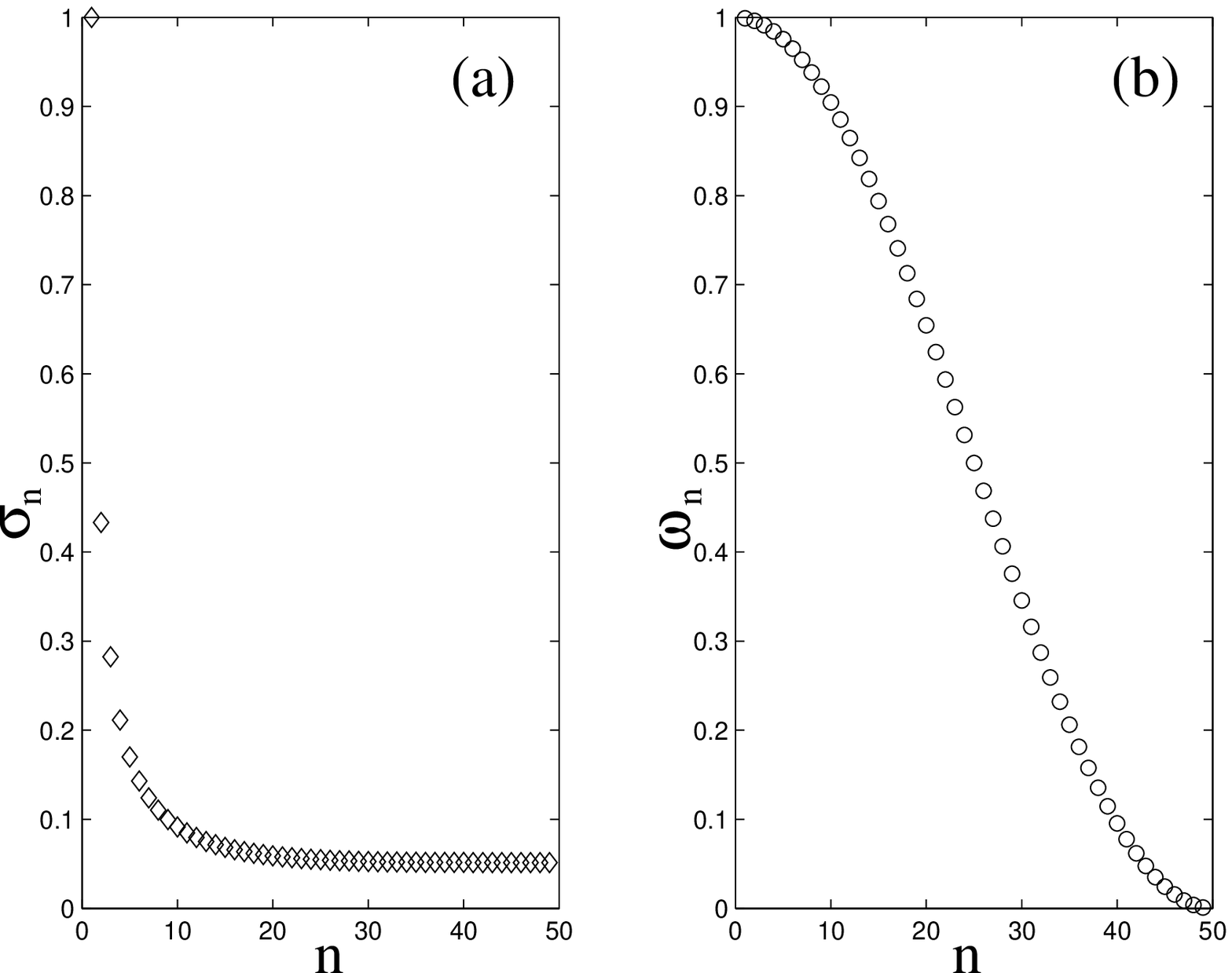}}
\caption{ {\small (a) A plot of the ordered Hankel singular values (HSV) for the homogeneous, harmonic oscillator
chain. The spring constants are uniformly taken to be $k=0.25$. 
The HSV are plotted for $T \propto N^{2}$ where $N = 49$.  The distribution of HSV 
remains essentially unchanged for any larger choice of $T$. (b) A plot of the 
frequencies for the same system.}}
\label{fig:uniform_lho}
\end{figure}

The bounds in equations \eqref{eq:ho_lb1} and \eqref{eq:ho_ub1}
also, at least for this model, provide information regarding finite size effects.
If we take the lattice spacing to be $b$ and the system size of the approximate system 
to be $L=qb$, the bounds then imply that (for $N \gg 1$)
\begin{equation}
\label{finite_size}
\displaystyle \lim_{T \to \infty} \frac{\| G -\tilde{G}_{2q}\|_{\mathcal{L}_2[0,T],i}}
{\| G \|_{\mathcal{L}_2[0,T],i}} = \mathcal{O}(L^{-1}).
\end{equation}
This result is not new, though these techniques provide a new way to obtain it.
Additionally, these techniques imply that for more general systems or experiments, reductions
may have quite a different dependence on the system size.  

It is apparent from  \eqref{eq:bal_real_main} that the balanced realization for the harmonic chain
weights driving of the momenta by $\mathbf{\Lambda}_{\Omega}^{-1/2}\mathbf{U}_d$.
Since $(\mathbf{\Lambda}_{\Omega})_{pp} = \omega_p \stackrel{N \gg 1}{\longrightarrow} \frac{\pi p}{2N}$,
this gives more weight to
momenta corresponding to small frequencies. 
Therefore it requires smaller gains to activate the slower modes. 
If driving gives
nontrivial initial conditions (i.e. impulses), 
this is equivalent to  slow-mode initial conditions being more easily
excited than fast-mode initial conditions.
While the balanced 
realization of the system implies that the internal states of the system are noninteracting, it 
also implies that 
differing normal modes  are not treated equally.  
This again suggests that the most natural coarse grainings are local coarse grainings.

Consider the latter conditions mentioned earlier, 
$\displaystyle 2\sqrt{\frac{k(N)}{m(N)}} \ge \frac{2(N+1)}{\pi}$ and $N \gg 1$. Here
$\omega_p>1$ for all $p \in \{1,\hdots,N\}$ and implies that $\alpha(p) = N+1-p$.
The upper and lower bounds for this case are respectively given by
 \begin{equation}
\label{eq:ho_lb2}
\begin{array}{c}
\| G  - \tilde{G}_{2q}\|_{\mathcal{L}_2[0,T],i} \ge 
 \frac{T}{2}\bigg( \cos\big(\frac{\pi(q+1)}{2(N+1)}\big)+ \Big(\cos\big(\frac{\pi(q+1)}{2(N+1)}\big)\Big)^{-1} \bigg) \\
= \frac{NT}{\pi}\Big(1 - \frac{\pi^2(q+1)^2}{4N^2}+\mathcal{O}\big(\left(\frac{q+1}{N}\right)^4\big)\Big),
\end{array}
\end{equation}
and
\begin{equation}
\label{eq:ho_ub2}
\begin{array}{c}
\| G  - \tilde{G}_{2q}\|_{\mathcal{L}_2[0,T],i} \le 
 \frac{eT}{2}\bigg( \cos\big(\frac{\pi(q+1)}{2(N+1)}\big)+ \Big(\cos\big(\frac{\pi(q+1)}{2(N+1)}\big)\Big)^{-1} \bigg) \\
= \frac{eNT}{\pi}\Big(1 - \frac{\pi^2(q+1)^2}{4N^2}+\mathcal{O}\big(\left(\frac{q+1}{N}\right)^4\big)\Big).
\end{array}
\end{equation}
This system is rather pathological
since any good approximant must have $q \propto N$, 
as seen in equation \eqref{eq:ho_lb2}.  It is impossible for the 
error to be made small unless $q$ is of the same order as $N$. That $q$ must scale 
as $N$ implies that this system does not admit the same nice reductions as the previous example.
Recall that for the previous example the relative error vanishes as $N \to \infty$ as 
long as $q \propto N^{\delta}$ for any $\delta $ such that $0<\delta<1$.  Thus, decent 
reductions must retain far more states than the previous example.      
Despite this pathology, 
the appropriate reductions project
 out the slower modes.  Since the same $\mathbf{U}_d$ may be used as before (i.e. essentially a Fourier transform),
 the coarse graining keeps the small distance behavior (fast modes) and projects
out the rest.  
For this system, high-frequency modes are more easily amplified,  
which explains the importance of including those modes 
in the approximate response. 

$\bullet$ {\bf The layered and random chains}

\begin{figure}
\centerline{\epsfxsize=6cm \epsfbox{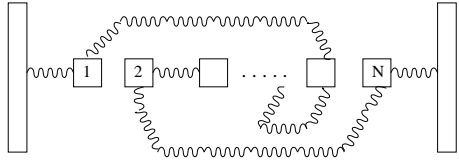}}
\caption{ {\small A spatially heterogeneous chain of linear oscillators. This is an example  
where spatially local coarse graining breaks down.}}
\label{fig:heterogeneous}
\end{figure}

When $\mathbf{U}_d$ is not a Fourier transform then these
amenable dispersion relations are not guaranteed.  Without such relations,
 modal reduction is not necessarily equivalent to local coarse graining.
For example, consider a system with uniform masses and spring constants 
that under a unitary change of basis has the same $\mathbf{\Omega}^2$ as 
in \eqref{eq:lin_osc_matrix} but without the spatial configuration 
of the linear chain 
(see Figure \ref{fig:heterogeneous}).  

This system is
 an oscillator system with nonlocal interactions and a  
spatial Fourier transform will not diagonalize $\mathbf{\Omega}$.
For this system, the appropriate reduction again would be a modal reduction 
since $\mathbf{B}$ and $\mathbf{C}$ only differ from $\mathbf{I}$ by a unitary transformation.
The exceptional thing about this example is that 
modal reduction  lumps together oscillators that are far apart 
yet directly connected to each other. 
Consequently, it is not a local coarse graining.
Although this system has the same characteristic frequencies and HSV as the 
homogeneous linear chain, 
long range interactions
disrupt the validity of local coarse graining.
Despite the artificial nature of this example, it illustrates the relationship between
heterogeneity, nonlocality, and long range interactions.  Frustration, induced by competing
interactions, also exemplifies these connections.

\begin{figure}
\centerline{\epsfxsize=6cm \epsfbox{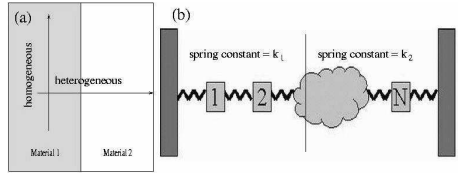}}
\caption{ {\small (a) A layered medium.  (b) A layered chain of linear oscillators.}}
\label{fig:layers}
\end{figure}

\begin{figure}
\centerline{\epsfxsize=6cm \epsfbox{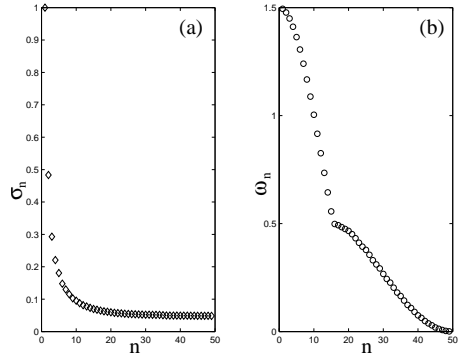}}
\caption{ { \small (a) A plot of the ordered HSV for the layered, harmonic oscillator
chain.  The HSV are plotted for $T \propto N^{2}$ where $N = 49$.  The spring constant on one side is 
taken to be $0.125$ one side and $0.375$ on the other. (b) A plot of the 
frequencies for the same system.}}
\label{fig:layered_lho}
\end{figure}

\begin{figure}
\centerline{\epsfxsize=6cm \epsfbox{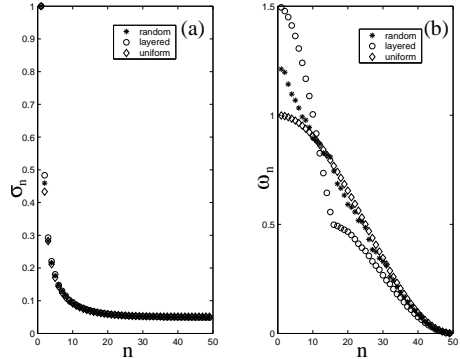}}
\caption{ { \small (a) A comparative plot of the ordered HSV for the different oscillator
chains.
The HSV are plotted for $T \propto N^{2}$ where $N = 49$.
 (b) Plots of the frequencies for the different oscillator chains.}}
\label{fig:combined}
\end{figure}

The scenario above is analogous to what happens in layered systems, as depicted in 
Figure \ref{fig:layers}. The material is homogeneous along one the vertical direction while it is 
heterogeneous in the other direction (transverse direction).  The importance of such
examples cannot be overemphasized, for they imply 
that the common practice
of local coarse graining will not always apply to heterogeneous systems
\footnote{Under reasonably restrictive assumptions about the nature of there heterogeneities, the theory of homogenization allows one to  effectively locally coarse grain an inhomogeneous system.}.  
Figure \ref{fig:layered_lho}(a) depicts the HSV for a one dimensional layered harmonic chain, a
variant of the scenario depicted in Figure \ref{fig:layers}.  
 The spring constants are 
taken to be $0.125$ on one side and $0.375$ on the other.
The HSV are distributed similarly to those of the homogeneous chain, as 
seen by comparing Figures \ref{fig:uniform_lho}(a) and \ref{fig:layered_lho}(a) 
or by inspection of Figure \ref{fig:combined}(a).
However, upon comparing Figures \ref{fig:uniform_lho}(b) and \ref{fig:layered_lho}(b),
the frequencies are quite different.  Additionally,
the transformation that diagonalizes $\Omega$ does not behave like a Fourier transform. 
Consequently, the appropriate coarse graining for this system is nonlocal. 

When the spring constants are uniformly, randomly sampled from the 
interval $[0.125,0.375]$, somewhat surprisingly, the HSV are distributed 
almost identically to those of the homogeneous chain.
As in the case of the layered chain, the frequencies of the random chain are different 
from those of the homogeneous one.
Just as in the previous case, the coarse grainings for this system are nonlocal.
In each case, when there is greater variation in the values of $k$, the distribution of HSV
start to vary more from the purely homogeneous case. 

\begin{figure}
\centerline{\epsfxsize=6cm \epsfbox{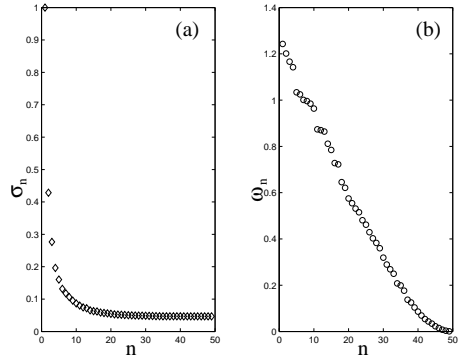}}
\caption{ {\small (a) A plot of the ordered HSV for the harmonic oscillator
chain with uniformly sampled random springs.  The spring constants are sampled 
from the interval  $[0.125,0.375]$.  The HSV are plotted for $T \propto N^{2}$ where $N = 49$.
 (b) A plot of the frequencies for the same system.}}
\label{fig:random_lho}
\end{figure}

\section{Conclusions and Future Directions}
\label{sec:conclusion}

Attempting to approximate systems by  only considering their relevant details
is not new in physics.  For instance, resumming  Feynman diagrams 
corresponding to prevalent physical processes attempts to do this.
Such methods fail since they are not systematic.
This is the novelty of this work. Not only does it provide a systematic way of determining 
how to coarse grain an arbitrary (linear) system, it also establishes how to implement 
the coarse graining in an
algorithmic fashion.  Furthermore, it is complementary to both the renormalization group (RG) 
\cite{chen96,oono00,nozaki01} and the projection-operator formalism because it
removes much of the  ambiguity in coarse graining.
Additionally, the Hankel methods can be used for arbitrary linear systems, not simply for generalized
Hamiltonian systems.  However, for general systems, the gramians are not guaranteed to take such 
simple forms.  

This paper also answers an important question that was raised by 
Hartle and Brun \cite{brun99}.  
In that work, both the quantum and classical aspects of the harmonic chain (on a ring) were investigated.
In particular, connections were made between different coarse grainings and the determinacy of the coarsened
equations of motion (in the classical case) and also between different coarse grainings and decoherence 
 (in the quantum case).  They conjectured  that local coarse graining led to more deterministic 
equations of motion for the coarsened degrees of freedom than nonlocal coarse graining, at least 
when the fast-mode initial conditions are thermally distributed.  
The induced noise for the locally-coarsened description was less than that for the nonlocal one.  However, as
was acknowledged in their work, the coarse grainings that were investigated constituted a set of 
measure zero of all the possible coarse grainings.  
This paper extends that work by considering arbitrary coarse grainings and more general quadratic 
Hamiltonians than just the harmonic chain.  For the homogeneous harmonic chain with 
$\displaystyle 2\sqrt{\frac{k}{m}} \le 1$, we confirm  
 that local coarse graining 
induces less noise than nonlocal coarse graining.
Additionally, we establish that this conclusion does not generalize to arbitrary oscillator systems. 
In fact, we have shown that this relationship is contingent on the dispersion relation and 
how the HSV depend on the 
normal-mode frequencies.  Consequently, for a different oscillator system, it is 
possible that nonlocal coarse grainings will yield the most deterministic equations of motion.

There are many theoretical directions in which this work can be taken, however,
the greatest pool of problems are those that relate to physically-motivated models.
The methods introduced here should be quite useful when  applied  to any number of physical systems 
where local coarse graining fails.  
For instance, inhomogeneous systems like layered or disordered systems
are prime nontrivial candidates.
Additionally, this work is ideally suited  for nonequilibrium systems.  In particular,
since it identifies the degrees of freedom that seem both most \lq\lq excitable'' and
\lq\lq observable'', it may be appropriate for revealing the true nature of effective temperatures \cite{cug97}.
For granular systems, this would be a big step towards 
identifying the importance of 
such mysterious quantities as the granular temperature and the free volume.   
Accordingly, it is in these directions, among others, that future work  using these methods should be taken.

 
\section*{Acknowledgments}
The author thanks Professors J. Hartle and I. Mezic for the 
conversations that culminated into this paper.  Special thanks 
are also due to C. Maloney for his many questions and criticisms
of this work.  This work has also benefited from 
comments by Professor D. Cai.
Lastly, the encouragements and suggestions from K. Reynolds and Professor 
J. Carlson are greatly appreciated and have been crucial to the completion of 
this work.
This work was supported by the David and Lucile Packard Foundation, NSF Grant No. DMR-9813752,
and EPRI/DoD through the Program on Interactive Complex Networks.


\appendix

\section{Infinite time horizon results for finite time horizon problems}
\label{sec:infiniteH}

Approximating conservative linear systems over an infinite time horizon inevitably 
leads to divergences.  This may be understood from the fact that the gramians become
unbounded due to the infinite time horizon.
A standard way to regulate this divergence is by approximating the system 
over a finite time horizon. Alternatively, the system can 
be exponentially discounted and considered over an infinite time horizon.

In this appendix we express the upper and lower bounds 
for the approximation of the input-output operator over a finite time 
horizon in terms of exponentially-discounted infinite-time-horizon Hankel singular values.
Although this analysis is only applied to conservative systems, 
we find the  bounds for arbitrary finite dimensional systems 
that admit LTI, causal realizations.   
Given a system realization $(\mathbf{A},\mathbf{B},\mathbf{C})$, 
we denote the input-output operator and its 
order $r$ approximant  respectively by $G$ and $\tilde{G}_r$. Similarly, for the 
exponentially discounted system (i.e. with  system matrix $-a\mathbf{I}+\mathbf{A}$), 
the input-output operator and its 
approximant are denoted by $G^{(a)}$ and $\tilde{G}^{(a)}_r$, respectively.
Additionally, the finite time horizon HSV are given by 
$\sigma_1 \ge \sigma_2 \ge \hdots \ge \sigma_n$ while the infinite time horizon singular values 
are given by $\sigma_1(a) \ge \sigma_2(a) \ge \hdots \ge \sigma_n(a)$.

Equation \eqref{eq:hinfin_lb} as it is stated is equally valid for infinite or finite time horizons.
However, we intend to relate $\| G-\tilde{G}_r \|_{\mathcal{L}_2[0,T],i}$ 
to the singular values $\{\sigma_i(a): 1 \le i \le n\}$.  The following new theorem establishes
the relation of interest.
\begin{theorem}[Lower Bound]
Given a LTI, causal system $G$ with $n$ dimensional minimal realization $(\mathbf{A},\mathbf{B},\mathbf{C})$.
If there exists an \lq\lq$a$'' such that $-a\mathbf{I}+\mathbf{A}$ is a stable system matrix then 
for any order $r$ (or less) approximant $\tilde{G}_r$
\begin{displaymath} 
\displaystyle \| G-\tilde{G}_r \|_{\mathcal{L}_2[0,T],i} \ge 
\big(1-e^{-2aT}\|e^{\mathbf{A}^{\dagger}T}e^{\mathbf{A}T}\|_{\mathbb{C}^n,i}\big)\sigma_{r+1}(a)
\end{displaymath}
\end{theorem}
\emph{Proof}:  

Since the Hankel operator is simply a projection of the original input output
operator we initially trivially find for an arbitrary $\tilde{G}_r$,
\begin{equation}
\displaystyle \| G-\tilde{G}_r \|_{\mathcal{L}_2[0,T],i} \ge 
\| \Gamma_{G}-\Gamma_{\tilde{G}_r} \|_{\mathcal{L}_2[0,T],i}
\ge \| \Gamma_{G}-\mathcal{K}_r \|_{\mathcal{L}_2[0,T],i},
\end{equation}
where $\mathcal{K}_r$ is an arbitrary rank $r$ operator that is not restricted to 
be of Hankel form.  The last inequality arises because it is 
known that $\| \Gamma_{G}-\Gamma_{\tilde{G}_r} \|_{\mathcal{L}_2[0,T],i} \ge \sigma(\Gamma_G)$
and equality is not always possible since $\Gamma_{\tilde{G}_r}$ is of Hankel form.  
However, equality is achievable for an arbitrary operator $\mathcal{K}_r$.

The primary nontrivial step in this proof requires the observation that 
the each of the eigenvalues of the balanced gramian, $\bar{W}^{(a)}$, associated to 
$\Gamma_{G^{(a)}}= \Gamma^{(a)}$ are decreasing functions of $a$.
This can be seen by noting that for any vector $\zeta \in \mathbb{R}^n$ and for $b<a$,
\begin{equation}
\left<\zeta,\left(\bar{W}^{(b)}-\bar{W}^{(a)}\right)\zeta\right> \ge 0.
\end{equation}
This means that $\sigma_k(a) \le \sigma_k(b)$ for $a>b$.
From this observation it is then follows that
\begin{equation}
\| \Gamma_{G}-\mathcal{K}_r \|_{\mathcal{L}_2[0,T],i} \ge
\| \Gamma^{(a)}-\widetilde{\mathcal{K}}_r \|_{\mathcal{L}_2[0,T],i}.
\end{equation}

Then using that for any two bounded linear operators, $\mathcal{A}$ and $\mathcal{B}$,
in the induced norm
$\|\mathcal{A}\mathcal{B}\| \le \|\mathcal{A}\|\|\mathcal{B}\|$ and that 
\begin{equation}
\|e^{\mathbf{A}T}\|_{\mathbb{C}^n,i} = 
\|e^{\mathbf{A}^{\dagger}T}e^{\mathbf{A}T}\|_{\mathbb{C}^n,i}^{1/2},
\end{equation}
we finally obtain
\begin{equation}
\begin{array}{c}
\| \Gamma^{(a)}-\widetilde{\mathcal{K}}_r \|_{\mathcal{L}_2[0,T],i} \ge
\| \Gamma^{(a)}-\widetilde{\mathcal{K}}_r \|_{\mathcal{L}_2[0,\infty],i}-
\| \Gamma^{(a)}-\widetilde{\mathcal{K}}_r \|_{\mathcal{L}_2[T,\infty],i}
\\
\ge 
\big(1-e^{-2aT}\|e^{\mathbf{A}^{\dagger}T}e^{\mathbf{A}T}\|_{\mathbb{C}^n,i}\big)\sigma_{r+1}(a).
\end{array}
\end{equation}
Hence we arrive at the desired result,
\begin{equation} 
\displaystyle \| G-\tilde{G}_r \|_{\mathcal{L}_2[0,T],i} \ge 
\big(1-e^{-2aT}\|e^{\mathbf{A}^{\dagger}T}e^{\mathbf{A}T}\|_{\mathbb{C}^n,i}\big)\sigma_{r+1}(a).
\end{equation}
In the case where $\mathbf{A}$ 
has no Jordan blocks the above result simplifies to
\begin{equation}
\displaystyle \| G-\tilde{G}_r \|_{\mathcal{L}_2[0,T],i} \ge 
\big(1-e^{2T \max \, \mathrm{Re} \, \lambda(-a\mathbf{I}+\mathbf{A})}\big)\sigma_{r+1}(a).
\end{equation} 
\begin{flushright} 
$\square$
\end{flushright}

This lower bound is completely consistent
with \eqref{eq:hinfin_lb} for stable systems.  
Although we only use the upper bound in determining the relative error, we include it in 
the following.
Recall from equation \eqref{eq:up_bound} that
the upper bound is only valid for stable systems (for infinite time).  In fact, 
it is almost exclusively proven for stable systems 
in the literature.
The following upper bounds, valid for finite time horizon, are taken from \cite{sznaier02}. 
\begin{theorem}[Upper Bound]
\label{ith_ub}
Given a LTI, causal system $G$ with $n$ dimensional minimal realization $(\mathbf{A},\mathbf{B},\mathbf{C})$,
if \lq\lq$a$'' is such that $-a\mathbf{I}+\mathbf{A}$ is a stable system matrix 
and $\| G^{(a)} \|_{\mathcal{L}_2,i} < \gamma$, then:
\begin{displaymath} 
\| G \|_{\mathcal{L}_2[0,T],i} < \gamma e^{aT}
\end{displaymath}
\end{theorem}
\emph{Proof}: This results follows from the differential version of the bounded real lemma.
For details refer to \cite{sznaier02}.  \begin{flushright}$\square$\end{flushright}

We already know from equation \eqref{eq:up_bound} how to obtain an upper bound for
the approximation error, $\| G^{(a)} - \tilde{G}^{(a)}_r \|_{\mathcal{L}_2,i}$. 
We combine this fact with Theorem \ref{ith_ub} to obtain the following corollary.
\begin{corollary}[Upper Bound]
Given a LTI, causal system $G$ with $n$ dimensional minimal realization $(\mathbf{A},\mathbf{B},\mathbf{C})$,
if \lq\lq$a$'' is such that $-a\mathbf{I}+\mathbf{A}$ is a stable system matrix 
then there exists an order $r$ input-output operator, $\tilde{G}_r$ obtained by 
balanced truncation such that
\begin{displaymath} 
\| G  - \tilde{G}_r\|_{\mathcal{L}_2[0,T],i} < 2e^{aT}\sum_{j=1}^k \sigma^{\mathrm{dist}}_{i_j}(a),
\end{displaymath}
where $\sigma^{\mathrm{dist}}_{i_j}(a), 1 \le j \le k$ are the distinct infinite time horizon
HSV from the set $\{\sigma_{r+1}(a), \hdots ,\sigma_n(a)\}$. 
\end{corollary}
To be precise, an algorithm to obtain $\tilde{G}_r$ is as follows.  First find the realization for
 $\tilde{G}^{(a)}_r$ by truncating the balanced realization of $(-a\mathbf{I}+\mathbf{A},\mathbf{B},\mathbf{C})$.
Denote the resulting realization by  $(\mathbf{A}_r,\mathbf{B}_r,\mathbf{C}_r)$.  A realization
for  $\tilde{G}_r$ is then just  $(a\mathbf{I}+\mathbf{A}_r,\mathbf{B}_r,\mathbf{C}_r)$.

\section{Calculation of the Gramians}
\label{sec:calculation}

In this appendix we intend 
to calculate the balanced form of the gramians for oscillator systems
(i.e. of systems of the form found in equation \eqref{eq:gen_ham}).  This 
entails calculating the damped (exponentially discounted) gramians:
\begin{equation}
\label{eq:damp_gram}
\begin{array}{c}
\mathbf{W}_c^{(a)} = \int_0^{\infty} e^{-2at}e^{\mathbf{A}t}e^{\mathbf{A}^{\dagger}t} \mathrm{d}t \\
\mathbf{W}_o^{(a)} = \int_0^{\infty} e^{-2at}e^{\mathbf{A}^{\dagger}t}e^{\mathbf{A}t} \mathrm{d}t
\end{array}
\end{equation}

However, first let us introduce the following notations and conventions.
Recall that any matrix, $\mathbf{S}$, may be expressed in terms of the canonical matrix units,
$\mathbf{e}_{ij}$.  In other words, 
\begin{displaymath}
\mathbf{S} = \sum_{i,j} S_{ij}\mathbf{e}_{ij}
\end{displaymath}
where each $S_{ij}$ is just a complex number.  
For instance, in the case of $2\times2$ matrices,  
\begin{displaymath}
\mathbf{e}_{12} = \left[ \begin{array}{cc} 0 & 1 \\ 0 & 0 \end{array} \right] 
\end{displaymath}
Additionally, for this section, 
$\mathbf{Q} = \left[ \begin{array}{cc} 0 & 1 \\ -1 & 0 \end{array} \right]$.
Lastly, we frequently use the algebraic tensor product, $\otimes$ \footnote{Often referred to as the dyadic product.}.  
For instance, suppose 
\begin{displaymath}
\mathbf{A} = \left[ \begin{array}{cc} A_{11} & A_{12} \\ 
A_{21} & A_{22} \end{array} \right]
\end{displaymath}
then
\begin{displaymath}
\mathbf{A}\otimes\mathbf{B} = \left[ \begin{array}{cc} A_{11}\mathbf{B} & A_{12}\mathbf{B} \\ 
A_{21}\mathbf{B} & A_{22}\mathbf{B} \end{array} \right]
\end{displaymath}

First note that if we define 
\begin{equation}
\label{eq:Rmatrix}
\mathbf{R} = \mathbf{e}_{11}\otimes\mathbf{\Omega}^{-1/2} + \mathbf{e}_{22}\otimes\mathbf{\Omega}^{1/2}
\end{equation}
then easily it follows that
\begin{equation}
\label{eq:transform1}
\begin{array}{c}
\mathbf{A} = 
 \mathbf{e}_{12}\otimes\mathbf{I} - \mathbf{e}_{21}\otimes\mathbf{\Omega}^2 =
\left[ \begin{array}{cc} 
0 &\mathbf{I} \\  - \mathbf{\Omega}^2 & 0 
\end{array} \right] \\
 \stackrel{\mathbf{R}}{\longrightarrow} 
\mathbf{R}^{-1}\mathbf{A}\mathbf{R} = \mathbf{Q}\otimes\mathbf{\Omega}
= \left[ \begin{array}{cc} 
0 &\mathbf{\Omega} \\  - \mathbf{\Omega} & 0 
\end{array} \right]
\end{array}
\end{equation}
From this, one then finds that 
\begin{equation}
\label{eq:trans_cgram}
\begin{array}{c}
\mathbf{W}_c^{(a)} = 
\mathbf{R}\int_0^{\infty} e^{-2at}e^{\mathbf{Q}\otimes\mathbf{\Omega}t}
\mathbf{R}^{-1}\mathbf{R}^{-1}e^{-\mathbf{Q}\otimes\mathbf{\Omega}t} \mathrm{d}t\mathbf{R} \\
= \mathbf{R}\int_0^{\infty} e^{-2at}e^{\mathbf{Q}\otimes\mathbf{\Omega}t}
\left[ \begin{array}{cc} 
\mathbf{\Omega} & 0 \\  0 & \mathbf{\Omega}^{-1} 
\end{array} \right]
e^{-\mathbf{Q}\otimes\mathbf{\Omega}t} \mathrm{d}t\mathbf{R}
\end{array}
\end{equation}
However, using that $e^{\mathbf{Q}\otimes\mathbf{\Omega}t}$ $= 
\mathbf{I}\otimes\cos\mathbf{\Omega}t + \mathbf{Q}\otimes\sin\mathbf{\Omega}$
we finally arrive at
\begin{equation}
\label{eq:cgram_result}
\begin{array}{c}
\mathbf{W}_c^{(a)} = \mathbf{R}\int_0^{\infty} e^{-2at}\times \\
\left[\begin{array}{cc}
\mathbf{\Omega}\cos^2\mathbf{\Omega}t + \mathbf{\Omega}^{-1}\sin^2\mathbf{\Omega}t &
\frac{1}{2}\mathbf{\Omega}^{-1}(\mathbf{\Omega}^{-1}-\mathbf{\Omega})\frac{\mathrm{d}}{\mathrm{d}t}
\sin^2\mathbf{\Omega}t \\
\frac{1}{2}\mathbf{\Omega}^{-1}(\mathbf{\Omega}^{-1}-\mathbf{\Omega})\frac{\mathrm{d}}{\mathrm{d}t}
\sin^2\mathbf{\Omega}t &
\mathbf{\Omega}^{-1}\cos^2\mathbf{\Omega}t + \mathbf{\Omega}\sin^2\mathbf{\Omega}t
\end{array}\right]\mathrm{d}t \ \mathbf{R}
\\
= \frac{1}{4a}\mathbf{R}\left[\begin{array}{cc}
\mathbf{\Omega} + \mathbf{\Omega}^{-1} & 0 \\
0 & \mathbf{\Omega} + \mathbf{\Omega}^{-1}
\end{array}\right]\mathbf{R} + 
\frac{1}{4a}\mathbf{R}\times \\ 
\left[\begin{array}{cc}
-a^2\mathbf{\Omega}^{-1}(\mathbf{I}-\mathbf{\Omega}^2)
(a^2\mathbf{I}+\mathbf{\Omega}^2)^{-1} &
a(\mathbf{I}-\mathbf{\Omega}^2)(a^2\mathbf{I}+\mathbf{\Omega}^2)^{-1} \\
a(\mathbf{I}-\mathbf{\Omega}^2)(a^2\mathbf{I}+\mathbf{\Omega}^2)^{-1} &
 a^2\mathbf{\Omega}^{-1}(\mathbf{I}-\mathbf{\Omega}^2)
(a^2\mathbf{I}+\mathbf{\Omega}^2)^{-1}
\end{array}\right]\mathbf{R}. \\
\end{array}
\end{equation}
Similarly for the observability gramian we obtain
\begin{equation}
\label{eq:ogram_result}
\begin{array}{c}
\mathbf{W}_o^{(a)} =
 \frac{1}{4a}\mathbf{R}^{-1}\left[\begin{array}{cc}
\mathbf{\Omega} + \mathbf{\Omega}^{-1} & 0 \\
0 & \mathbf{\Omega} + \mathbf{\Omega}^{-1} 
\end{array}\right]\mathbf{R}^{-1} + 
 \frac{1}{4a}\mathbf{R}^{-1}\times \\ 
\left[\begin{array}{cc}
 a^2\mathbf{\Omega}^{-1}(\mathbf{I}-\mathbf{\Omega}^2)
(a^2\mathbf{I}+\mathbf{\Omega}^2)^{-1} &
a(\mathbf{I}-\mathbf{\Omega}^2)(a^2\mathbf{I}+\mathbf{\Omega}^2)^{-1} \\
a(\mathbf{I}-\mathbf{\Omega}^2)(a^2\mathbf{I}+\mathbf{\Omega}^2)^{-1} &
-a^2\mathbf{\Omega}^{-1}(\mathbf{I}-\mathbf{\Omega}^2)
(a^2\mathbf{I}+\mathbf{\Omega}^2)^{-1}
\end{array}\right]\mathbf{R}^{-1}.
\end{array}
\end{equation}

From equations \eqref{eq:cgram_result} and \ref{eq:ogram_result} it 
follows after using $\mathbf{U}_{d}$ to diagonalize $\mathbf{\Omega}$
and taking the small \lq\lq$a$'' limit that the balanced gramian, without ordered eigenvalues,
is given by
\begin{displaymath}
\bar{\mathbf{W}}_c = \bar{\mathbf{W}}_o = 
 \frac{1}{4a}\left[\begin{array}{cc}
\mathbf{\Lambda}_{\Omega} + \mathbf{\Lambda}_{\Omega}^{-1} + \mathcal{O}(a^2) & \mathcal{O}(a) \\ 
\mathcal{O}(a) & \mathbf{\Lambda}_{\Omega} + \mathbf{\Lambda}_{\Omega}^{-1} + \mathcal{O}(a^2)  \end{array}\right].
\end{displaymath}

\section{Restrictions on the time horizon/exponential discounting}
\label{sec:restrictions}
\subsection{Relevant matrix perturbation}

We start with a matrix $\mathbf{L_{\epsilon}}$ of the form,

\begin{equation}
\label{eq:pert1}
\mathbf{L}_{\epsilon} =
\left( 
\begin{array}{cc}
\mathbf{M} & \epsilon \mathbf{N} \\
\epsilon \mathbf{N} & \mathbf{M} 
\end{array}
\right).
\end{equation}

The question we intend to address is, how do the $\epsilon$ terms perturb the spectrum of  
 $\mathbf{L}_{0}$.  

\begin{equation}
\label{eq:pert2}
\begin{array}{cc}
\mathrm{det}(\lambda \mathbf{I} - \mathbf{L}_{\epsilon})
= \mathrm{det} 
\left(
\begin{array}{cc}
\lambda \mathbf{I} - \mathbf{M} & -\epsilon \mathbf{N} \\
-\epsilon \mathbf{N} & \lambda \mathbf{I} - \mathbf{M} 
\end{array}
\right)   \\
=
\mathrm{det} 
\left(
\begin{array}{cc}
\lambda \mathbf{I} - \mathbf{M} & -\epsilon \mathbf{N} \\
0 & \lambda \mathbf{I} - \mathbf{M}-\epsilon^2 N (\lambda \mathbf{I}-M)^{-1} N
\end{array}
\right) 
\\
=
\mathrm{det}\big( \lambda \mathbf{I} - \mathbf{M} \big) \mathrm{det} \big( \lambda \mathbf{I} -
 \mathbf{M} - \epsilon^2 N (\lambda \mathbf{I}-M)^{-1} \mathbf{N} \big) \\
=\big( \mathrm{det} (\lambda \mathbf{I} - \mathbf{M}) \big)^2 \mathrm{det} \Big( \mathbf{I} - 
\epsilon^2 \big( (\lambda \mathbf{I}-\mathbf{M})^{-1} \mathbf{N} \big)^2 \Big)
\end{array}
\end{equation}

However, recall that given an invertible matrix $\mathbf{P}$,
\begin{equation}
\label{eq:inverse}
 \mathbf{P}^{-1} = \big(\mathrm{det}(\mathbf{P}) \big)^{-1} \mathrm{adj}(\mathbf{P}),
\end{equation}
where $\mathrm{adj}$ is
the formal matrix adjoint.
Combining \eqref{eq:inverse} with the fact that $\epsilon \ll 1$, we obtain:
\begin{equation}
\label{eq:pert3}
\mathrm{det}(\lambda \mathbf{I} - \mathbf{L}_{\epsilon}) \approx
\Big( \mathrm{det}(\lambda \mathbf{I}-\mathbf{M}) \Big)^2 
- \epsilon^2 \mathrm{Tr} \Big( \big(\mathrm{adj}(\lambda \mathbf{I}-\mathbf{M}) \mathbf{N}\big)^2 \Big).
\end{equation}
Now, let us evaluate the above at an eigenvalue of $\mathrm{L}_{\epsilon}$ that is 
``near'' an eigenvalue of $\mathrm{L}_{0}$.  Thus, $\lambda = \lambda_0 +\hat{\lambda}(\epsilon)$,
where $\lambda_0$ is an eigenvalue of $\mathrm{L_{0}}$ and 
$\hat{\lambda}(\epsilon) \to 0$ as $\epsilon \to 0$.
First consider the
transformation $\mathbf{U}$ that has the property that
\begin{equation}
\label{eq:diag}
\mathbf{U}^{-1} \mathbf{M} \mathbf{U} = 
\left( 
\begin{array}{cc}
\tilde{\mathbf{M}} & 0 \\
0 & \lambda_0 \mathbf{I}_{r\times r}
\end{array}
\right),
\end{equation}
where $r \ge 1$.
From this we find
\begin{equation}
\label{eq:pert4}
\begin{array}{l}
\mathrm{det}(\lambda_0 \mathbf{I} - \mathbf{M} +\hat{\lambda}(\epsilon) \mathbf{I} )
= \hat{\lambda}^{r}(\epsilon) \mathrm{det}(\lambda_0 \mathbf{I}-\tilde{\mathbf{M}}) 
\mathrm{det} \big(\mathbf{I} + \hat{\lambda}(\epsilon)(\lambda_0 
\mathbf{I} - \tilde{\mathbf{M}})^{-1}\big)  \\
=\hat{\lambda}^{r}(\epsilon) \mathrm{det}(\lambda_0 \mathbf{I}-\tilde{\mathbf{M}})
\Big[ 1 + \hat{\lambda}(\epsilon)\big(\mathrm{det}(\lambda_0 
\mathbf{I}-\tilde{\mathbf{M}})\big)^{-1} \mathrm{Tr}\big( \mathrm{adj}
(\lambda^* \mathbf{I} - \tilde{\mathbf{M}}) \big) \Big] \\
+ \mathcal{O}(\hat{\lambda}^{r+2}(\epsilon))
= \hat{\lambda}^{r}(\epsilon) \mathrm{det} (\lambda_0 \mathbf{I}-\tilde{\mathbf{M}})
+  \hat{\lambda}^{r+1}(\epsilon) \mathrm{Tr}\big( \mathrm{adj}
(\lambda_0 \mathbf{I} - \tilde{\mathbf{M}}) \big) \\
+ \mathcal{O}(\hat{\lambda}^{r+2}(\epsilon)).
\end{array}
\end{equation}

Combining \eqref{eq:pert3} and \eqref{eq:pert4}, results in
\begin{equation}
\label{eq:pert5}
\begin{array}{cc}
\mathrm{det}(\lambda \mathbf{I} - \mathbf{L}_{\epsilon}) 
= \hat{\lambda}^{2r}(\epsilon) \big(\mathrm{det} (\lambda_0 \mathbf{I}-\tilde{\mathbf{M}})\big)^2 \\
- \epsilon^2 \mathrm{Tr} \Big( \big(\mathrm{adj}(\lambda_0 \mathbf{I}-\mathbf{M}) \mathbf{N}\big)^2 \Big)
+\mathcal{O}(\mathrm{max} \{ \hat{\lambda}^{2r+1}(\epsilon), \epsilon^2 \hat{\lambda}(\epsilon) \}) = 0.
\end{array}
\end{equation}
This naively suggests that $|\hat{\lambda}(\epsilon)| \sim \epsilon^{1/r} $.  However, unless 
$\mathbf{M}$ has Jordan blocks, $\mathrm{adj}(\lambda \mathbf{I} - \mathbf{M})\big|_{\lambda=\lambda_0}
= 0$.  
We will consider $\mathbf{M}$ to be diagonalizable, in order to discern the  behavior of 
$\hat{\lambda}(\epsilon)$.  Thus, in the basis where $\mathbf{M} \to \mathbf{\Lambda_M}$ (i.e. the basis
where $\mathbf{M}$ is diagonal), the following holds:
\begin{equation}
\label{eq:matrix1}
\begin{array}{ll}
\displaystyle \frac{d^p}{d\lambda^p} \mathrm{adj}(\lambda \mathbf{I} - \mathbf{\Lambda_M})\big|_{\lambda=\lambda_0}
 = 0 &  \textrm{For all $p < r-1$}.
\end{array}
\end{equation}
In this case, we find
\begin{equation}
\label{eq:matrix2}
\begin{array}{cc}
\displaystyle \mathrm{adj}(\lambda \mathbf{I} - \mathbf{\Lambda_M})\big|_{\lambda=\lambda_0+\hat{\lambda}(\epsilon)} =
\frac{\hat{\lambda}^{r-1}(\epsilon)}{(r-1)!}
\frac{d^{r-1}}{d\lambda^{r-1}} \mathrm{adj}(\lambda \mathbf{I} - \mathbf{\Lambda_M}) + 
\mathcal{O}(\hat{\lambda}^{r}(\epsilon)) \\
\displaystyle = \frac{\hat{\lambda}^{r-1}(\epsilon)}{(r-1)!} \mathrm{det} (\lambda_0 \mathbf{I}-\tilde{\mathbf{M}})
\left( 
\begin{array}{cc}
0 & 0 \\
0 & \mathbf{I}_{r\times r}
\end{array}
\right) +\mathcal{O}(\hat{\lambda}^{r}(\epsilon)).
\end{array}
\end{equation}
   
Combining \eqref{eq:pert5} and \eqref{eq:matrix2} and provided that there exists a transformation
$\mathbf{V}$ such that $\mathbf{V}^{-1} \mathbf{M} \mathbf{V} = \mathbf{\Lambda_{M}}$ and 
 $\mathbf{V}^{-1} \mathbf{N} \mathbf{V} = \tilde{\mathbf{N}}$ we obtain the result
\begin{equation}
\label{eq:pert6}
\begin{array}{c}
\mathrm{det}(\lambda \mathbf{I} - \mathbf{L}_{\epsilon}) 
= \hat{\lambda}^{2r}(\epsilon) \big(\mathrm{det} (\lambda_0 \mathbf{I}-\tilde{\mathbf{M}})\big)^2
\displaystyle - \frac{1}{((r-1)!)^2}\epsilon^2 \hat{\lambda}^{2r-2}(\epsilon)\mathrm{Tr}(\tilde{\mathbf{N}}_{22})^2 \\
+ \mathcal{O}(\mathrm{max} \{ \hat{\lambda}^{2r+1}(\epsilon), \epsilon^2 \hat{\lambda}^{2r-1}(\epsilon) \}) = 0,
\end{array}
\end{equation}
where $\tilde{\mathbf{N}}_{22}$ is an $r \times r$ matrix that comes from
\begin{equation}
\label{eq:N_matrix}
\tilde{\mathbf{N}} = 
\left(
\begin{array}{cc}
\tilde{\mathbf{N}}_{11} & \tilde{\mathbf{N}}_{12} \\
\tilde{\mathbf{N}}_{21} & \tilde{\mathbf{N}}_{22}
\end{array}
\right).
\end{equation}
Hence we finally arrive at 
\begin{equation}
\label{eq:pert7}
|\hat{\lambda}(\epsilon)| = \frac{\epsilon}{(r-1)!} \sqrt{\big|\mathrm{Tr}(\tilde{\mathbf{N}}_{22})^2 
\big|} + \mathcal{O}(\epsilon^2).
\end{equation}

\subsection{Consequences:  Results put in context}
 
For $r=1$ and $\tilde{\mathbf{N}}$ is Hermitian, \eqref{eq:pert7} directly implies that the
 $i^{\mathrm{th}}$ eigenvalue of $\mathbf{M}$ is perturbed as 
\begin{equation}
\label{eq:pert8}
|\hat{\lambda}_i(\epsilon)| = \epsilon |\tilde{\mathrm{N}}_{ii}| + \mathcal{O}(\epsilon^2)
\le \epsilon \| \mathbf{N} \|_\infty + \mathcal{O}(\epsilon^2).
\end{equation}
This provides an upper bound on how the perturbation shifts the spectrum of the 
unperturbed matrix.

From Appendix \ref{sec:calculation}, we can make the following associations; $\epsilon = a$,
$\mathbf{M} = \mathbf{\Lambda}_{\Omega} + \mathbf{\Lambda}^{-1}_{\Omega}$, and
$\mathbf{N} = \mathbf{\Lambda}^{-2}_{\Omega}-\mathbf{I}$.
Our objective is to roughly determine the size of \lq\lq$a$'' such that the ordering of 
the Hankel singular values are preserved.  As in Section \ref{subsec:gen_osc},
given that $\lambda_j(\mathbf{\Omega}) \le \lambda_k(\mathbf{\Omega})$ for all $j<k$, let 
 $\alpha$ be a permutation such that 
 $\lambda_{\alpha(j)}(\mathbf{\Omega})+\lambda_{\alpha(j)}^{-1}(\mathbf{\Omega}) \le 
\lambda_{\alpha(k)}(\mathbf{\Omega})+\lambda_{\alpha(k)}^{-1}(\mathbf{\Omega})$ for all $k<j$.
Though it is somewhat conservative, the ordering about the $i^{th}$ unperturbed Hankel singular value 
is guaranteed provided that
\begin{equation}
\label{eq:local_ord}
\begin{array}{c}
a\left(| \lambda_{\alpha(i)}^{-2}-1| +| \lambda_{\alpha(i+1)}^{-2}-1|\right) \le
\lambda^{-1}_{\alpha(i)}+\lambda_{\alpha(i)}-\lambda^{-1}_{\alpha(i+1)}+\lambda_{\alpha(i+1)}, \\
 a\left(| \lambda_{\alpha(i-1)}^{-2}-1| +| \lambda_{\alpha(i)}^{-2}-1|\right) \le
\lambda^{-1}_{\alpha(i-1)}+\lambda_{\alpha(i-1)}-\lambda^{-1}_{\alpha(i)}+\lambda_{\alpha(i)}.
\end{array}
\end{equation}
It follows that the ordering of the HSV is guaranteed to be preserved provided that  
\begin{equation}
\label{eq:a_restriction}
a \le \min_{1\le i \le N-1} \frac{\lambda^{-1}_{\alpha(i)}+\lambda_{\alpha(i)}-\lambda^{-1}_{\alpha(i+1)}-
\lambda_{\alpha(i+1)}}
{| \lambda_{\alpha(i)}^{-2}-1| +| \lambda_{\alpha(i+1)}^{-2}-1|}.
\end{equation}


\begin{thebibliography}{}

\bibitem{sznaier02} M. Sznaier, A.C. Doherty, M. Barahona, J.C. Doyle, and H. Mabuchi,
\lq\lq A New Bound of the $\mathcal{L}_2[0,T]$-Induced Norm and Applications to Model Reduction,''
{\it Proceedings ACC} {\bf 2}, 1180 (2002).

\bibitem{barahona02} M. Barahona, A.C. Doherty, M. Sznaier, H. Mabuchi, and J.C. Doyle,
\lq\lq Finite Horizon Model Reduction and the Appearance of Dissipation of Hamiltonian Systems,'' 
{\it Proceedings of IEEE CDC} {\bf 4}, 4563 (2002).

\bibitem{reyn1} Reynolds, et. al. (in preparation).

\bibitem{rowley03} C. Rowley, (in preparation).

\bibitem{zwanzig73} R. Zwanzig,  \lq\lq
Nonlinear generalized langevin equations,''
{\it J. Stat. Phys.} {\bf 9}, 215-220 (1973).

\bibitem{mori65a} H. Mori,  \lq\lq
Transport, collective motion, and Brownian motion,''
{\it Prog. Theor. Phys.} {\bf 33}, 423-450 (1965).
\bibitem{mori65b} H. Mori,  \lq\lq
A continued-fraction representation of time-correlation functions,''
{\it Prog. Theor. Phys.} {\bf 34}, 399 (1965).


\bibitem{ford87} G.W. Ford and M. Kac,  \lq\lq
On the quantum langevin equation,''
{\it J. Stat. Phys.} {\bf 46}, 803-810 (1987).

\bibitem{gellman93} M. Gell-Mann and J. Hartle, \lq\lq
Classical equations for quantum systems,''
{\it Phys. Rev. D} {\bf 47}, 3345-3382 (1993).

\bibitem{brun99} T. Brun and J. Hartle, \lq\lq
Classical dynamics of the quantum harmonic chain,'' 
{\it Phys. Rev. D} {\bf 60}, 123503 (1999).


\bibitem{cug97} L.F. Cugliandolo, J. Kurchan, and L. Peliti,  \lq\lq
Energy flow, partial equilibration, and effective temperatures in systems with slow dynamics,''
{\it Phys. Rev. E} {\bf 55}, 3898-3914 (1997).
\bibitem{hansen} J.-P. Hansen and I.R. McDonald, {\it Theory of Simple Liquids}, 
2nd Edition (Academic Press,San Diego, 1990).
\bibitem{vankampen} N.G. van Kampen, {\it Stochastic Processes in Physics and Chemistry}, 
Revised and enlarged Edition (Elsevier, Amsterdam, 2001).
\bibitem{kubo} R. Kubo, M. Toda, N. Hashitsume, {\it Statistical Physics II}, 
2nd Edition (Springer-Verlag, Berlin, 1991).

\bibitem{glover84} K. Glover, \lq\lq
All optimal Hankel-norm approximations of linear-multivariable systems and their L infinity-error bounds,'' Int. J. Control {\bf 39}, 1115-1193 (1984).
\bibitem{feintuch} A. Feintuch, {\it Robust Control Theory in Hilbert Space},
 (Springer-Verlag, New York, 1998).
\bibitem{toeplitz1} A. B\"ottcher and B. Silbermann, 
{\it Introduction to Large Truncated Toeplitz Matrices}, (Springer-Verlag, New York, 1999).
\bibitem{toeplitz2} A. B\"ottcher and S.M. Grudsky, 
{\it Toeplitz Matrices, Asymptotic Linear Algebra, and Functional Analysis}, 
(Birkh\"auser, Boston, 2000).
\bibitem{toeplitz3} A. B\"ottcher and B. Silbermann, {\it Analysis of Toeplitz Operators},
(Springer-Verlag, New York, 1990).
\bibitem{robust1} G.E. Dullerud and F. Paganini, {\it A Course in Robust Control Theory}, 
(Springer-Verlag, New York, 2000).
\bibitem{peller} V.V. Peller, {\it Hankel Operators and thier Applications}, 
(Springer-Verlag, New York, 2003).
\bibitem{nikolski} N.K. Nikolski, {\it Operators, Functions and Systems:  An Easy Reading \ Volume I: 
Hardy, Hankel, and Toeplitz},(American Mathematical Society, Providence, 2002).
\bibitem{partington} J.R. Partington, {\it Interpolation, Identification, and Sampling},
(Oxford University Press, New York, 1997).


\bibitem{robust2} K. Zhou and J.C. Doyle, {\it Essentials of Robust Control}, (Prentice-Hall, New Jersey, 1998).

\bibitem{vinnicombe} G. Vinnicombe, {\it Uncertainty and Feedback:  $\mathcal{H}_{\infty}$ 
Loop-Shaping and the $\nu$-Gap Metric}, (Imperial College Press,Singapore,2000).

\bibitem{boyd} S. Boyd, L. El Ghaoui, E. Feron, and V. Balakrishnan, 
{\it Linear Matrix Inequalities in System and Control Theory},
(SIAM, Philadelphia, 1994).


\bibitem{scherpen93} J.M.A. Scherpen,  \lq\lq
Balancing for nonlinear-systems,''
{\it Syst. Contr. Lett.} {\bf 21}, 143-153 (1993).
\bibitem{scherpen96} J.M.A. Scherpen, \lq\lq
H-infinity balancing for nonlinear systems,''
{\it Int. J. Robust Nonlin. Contr.} {\bf 6}, 645-668 (1996). 
\bibitem{scherpen00} J.M.A. Scherpen,  \lq\lq
Minimality and local state decompositions of a nonlinear state space realization using energy functions,''
{\it IEEE Trans. Automat. Contr.} {\bf 45}, 2079-2086 (2000).

\bibitem{chen96} L.-Y. Chen, N. Goldenfeld, and Y. Oono,  \lq\lq
Renormalization group and singular perturbations: 
Multiple scales, boundary layers, and reductive perturbation theory,''
{\it Phys. Rev. E} {\bf 54},376-394 (1996).
\bibitem{oono00} Y. Oono, \lq\lq Renormalization and asymptotics,''
{\it Int. J. Mod. Phys. B} {\bf 14}, 1327-1361 (2000).
\bibitem{nozaki01} K. Nozaki and Y. Oono,  \lq\lq
Renormalization-group theoretical reduction,''
{\it Phys. Rev. E} {\bf 63}, 046101 (2001).
\bibitem{gold} N. Goldenfeld, {\it Lectures on Phase Transitions and the Renormalization Group},
(Perseus, Reading, MA, 1992).

\bibitem{Lesne} A. Lesne, {\it Renormalization Methods}, (John Wiley \& Sons, Chichester, UK, 1998). 

\bibitem{bricmont01} J. Bricmont, A. Kupiainen, and R. Lefevere, \lq\lq
Renormalizing the renormalization group pathologies,'' 
{\it Phys. Rep.} {\bf 348}, 5-31 (2001).
\bibitem{bricmont98} J. Bricmont, A. Kupiainen, and R. Lefevere,  \lq\lq
Renormalization group pathologies and the definition of Gibbs states,''
{\it Commun. Math. Phys.} {\bf 194}, 359-388 (1998).
\bibitem{bricmont88} J. Bricmont, A. Kupiainen, \lq\lq
Phase transition in the 3d random field Ising model,''
{\it Commun. Math. Phys.} {\bf 116}, 539-572 (1988).

\end{thebibliography}
\end{document}